\documentclass[aps,pra,reprint,superscriptaddress,letterpaper,twocolumn,longbibliography,floatfix]{revtex4-2}

\usepackage{polyglossia}
\setmainlanguage{english}

\usepackage{graphicx}
\usepackage[svgnames,dvipsnames,x11names]{xcolor}
\usepackage[colorlinks=true,linkcolor=SpringGreen4,citecolor=blue,urlcolor=Magenta3]{hyperref}

\usepackage{amsmath}
\usepackage{amssymb}
\usepackage{float}
\usepackage{amsthm}
\usepackage{amsfonts}
\usepackage{mathtools}
\usepackage{mathrsfs}
\usepackage{bm}
\usepackage{bbm}
\usepackage{empheq}
\usepackage{cases}
\usepackage{euscript}

\usepackage[draft,todonotes={textsize=scriptsize}]{changes}
\setuptodonotes{inline}
\definechangesauthor[name={Razvan Stanescu}, color=Blue4]{RS}
\definechangesauthor[name={Hugo Ribeiro}, color=SpringGreen4]{HR}

\newcommand{\bra}[1]{\langle #1|}
\newcommand{\ket}[1]{|#1\rangle}

\newcommand{\ketbra}[2]{\ket{#1}\!\bra{#2}}

\newcommand{\mm}[1]{\mathrm{#1}}
\newcommand{\abs}[1]{\left|#1\right|}

\newcommand{\di}[1]{\mathop{}\!\mathrm{d} #1}

\def \ud{\mathrm{d}}
\def \ue{\mathrm{e}}
\def \uf{\mathrm{f}}
\def \ug{\mathrm{g}}

\def \up{\mathrm{p}}

\def \uI{\mathrm{I}}

\def \uC{\mathrm{C}}
\def \uJ{\mathrm{J}}

\def \uT{\mathrm{T}}

\def \hsigma{\hat{\sigma}}

\def \hOmega{\hat{\Omega}}

\def \hb{\hat{b}}

\def \hn{\hat{n}}

\def \hH{\hat{H}}
\def \hV{\hat{V}}
\def \hU{\hat{U}}
\def \hO{\hat{O}}
\def \hT{\hat{T}}
\def \hS{\hat{S}}

\def \hW{\hat{W}}

\def \hP{\hat{P}}

\begin{document}

\title{Effective Hamiltonians for Predictive Quantum Control}

\author{Razvan Stanescu}
\affiliation{Department of Physics and Applied Physics, University of Massachusetts Lowell, Lowell, MA 01854, USA}
    
\author{Hugo Ribeiro}
\affiliation{Department of Physics and Applied Physics, University of Massachusetts Lowell, Lowell, MA 01854, USA}

\begin{abstract}
High-fidelity quantum control relies on accurate models of driven dynamics. We examine this requirement for single-qubit gates in
superconducting transmons by comparing control pulses derived from the standard Duffing approximation and from a Hamiltonian
constructed by diagonalizing the transmon eigenbasis. Using the same correction-pulse construction for both models, we show that
correction fields derived from the Duffing approximation can substantially reduce the gate error predicted by that model while
remaining less effective when combined with an independently calibrated baseline pulse in the diagonalized-transmon model. In the
fast-gate regime, such transferred corrections can even fail to improve over the uncorrected diagonalized-transmon baseline.  We
show that small model-dependent differences in both the energy spectrum and the representation of the drive operator can compound
during driven evolution, resulting in different predicted error generators and correction pulses. A mismatch in the accumulated AC
Stark phase provides one illustrative diagnostic of this dynamical model dependence.  We further demonstrate that the model
Hamiltonian informs the choice of control framework:~Omitting relevant leakage pathways or higher-order error channels can lead to
an overly restricted correction strategy. Including these channels motivates an extended correction framework that improves the
gate performance using the same physical control resources.
\end{abstract}

\maketitle

\section{Introduction}

The ability to steer quantum systems with high precision is a central requirement for quantum technologies. Across quantum
information processing, simulation, sensing, and metrology, useful dynamics are rarely obtained from the uncontrolled evolution of
a physical device alone. Instead, external fields must be engineered to implement target operations, suppress leakage from the
relevant Hilbert space, compensate unwanted dynamical phases, and reduce the effects of decoherence and control imperfections.
Precise control is particularly important in superconducting quantum processors, where microwave fields enable gate operations on
nanosecond timescales but can also excite transitions outside the intended computational subspace when the drives are strong or
broadband. For transmon qubits, the weak anharmonicity that suppresses charge-noise sensitivity also makes fast gates susceptible
to leakage and drive-induced phase errors~\cite{koch2007,schreier2008,blais2021,motzoi2009,gambetta2011,hyyppa2024}.

A large body of work has developed methods for finding control pulses that implement quantum operations with high fidelity.
Numerical quantum optimal-control methods, including GRAPE~\cite{khaneja2005}, Krotov-type methods~\cite{krotov1983},
chopped-random-basis approaches~\cite{doria2011}, and flexible gradient-based pulse-optimization techniques~\cite{machnes2018},
search over pulse shapes to optimize a fidelity functional subject to physical constraints. These methods have achieved
considerable success across platforms, including nuclear magnetic resonance, atoms, trapped ions, and superconducting
circuits~\cite{glaser2015,werschnik2007}. In superconducting qubits, numerical optimization can be especially powerful for short
gates, where perturbative pulse-shaping methods become less reliable. Its practical use, however, can be limited by the size of
the search space and by sensitivity to calibration errors, bandwidth constraints, pulse distortions, and discrepancies between the
simulated model and the physical device~\cite{machnes2018,genois2025,malarchick2026}.

Analytical and semi-analytical pulse-design methods provide a complementary route. For weakly anharmonic qubits, derivative
removal by adiabatic gate (DRAG) and related constructions use physical insight into the dominant leakage transition to suppress
population transfer outside the computational subspace and compensate drive-induced phase shifts~\cite{motzoi2009,gambetta2011}.
More recent approaches, including FAST DRAG and higher-derivative DRAG, shape the pulse spectrum to suppress spectral weight over
frequency intervals associated with leakage transitions and have enabled fast, low-leakage transmon gates without iterative
closed-loop optimization~\cite{hyyppa2024}. These methods reduce calibration overhead and provide interpretable design rules, but
their success depends on the assumptions used to identify the relevant transition frequencies, drive matrix elements, and error
channels.

Numerical and analytical control methods therefore share a common prerequisite:~The Hamiltonian used for pulse design must be
sufficiently accurate for the intended control task. Open-loop control is predictive only to the extent that the model captures
the driven dynamics of the device. A Hamiltonian that omits relevant levels, misrepresents transition frequencies or
control-operator matrix elements, or neglects important drive-induced effects can produce a pulse that performs well in simulation
but fails to realize the intended operation in a more faithful description of the system. This concern has motivated efforts to
incorporate hardware constraints, waveform distortions, robustness objectives, and data-driven model learning into quantum optimal
control~\cite{machnes2018,genois2025}.

Importantly, agreement at the level of static quantities does not by itself establish that an effective Hamiltonian is
sufficiently accurate for control design. Small differences in transition frequencies and control-operator matrix elements can
combine during driven evolution and produce quantitatively different dynamical error generators, even when the underlying static
approximations appear individually well justified. The model must therefore be assessed not only by how accurately it reproduces
the undriven spectrum, but also by whether it predicts the errors accumulated under the control protocol of interest. This
perspective is consistent with recent comparisons of DRAG and numerically optimized transmon gates, which show that the practical
advantage of numerical optimization depends on the gate-time regime, robustness requirements, and accuracy of the underlying
model~\cite{malarchick2026}.

The transmon provides a useful setting in which to investigate this model dependence because several effective Hamiltonians are
commonly used for control design. The circuit Hamiltonian contains the full cosine potential of the Josephson junction and is
specified by the charging and Josephson energies~\cite{koch2007,blais2021}. A widely used simplification is the Duffing
approximation, obtained by expanding the cosine potential to quartic order and treating the transmon as a weakly anharmonic
oscillator with constant adjacent-transition anharmonicity. This approximation provides much of the standard intuition underlying
leakage suppression and DRAG corrections. However, it modifies both the static spectrum and the representation of the physical
charge-drive operator. In the transmon eigenbasis, the drive couples through dressed charge matrix elements, whereas the Duffing
model replaces these elements with their harmonic-oscillator values. These differences may be small for weak and slow driving, but
they can become quantitatively important for fast gates, where leakage, AC Stark shifts, and other off-resonant effects increase
with drive strength~\cite{motzoi2009,gambetta2011,hyyppa2024}.

\begin{figure}[t]
    \includegraphics[width=0.99\columnwidth]{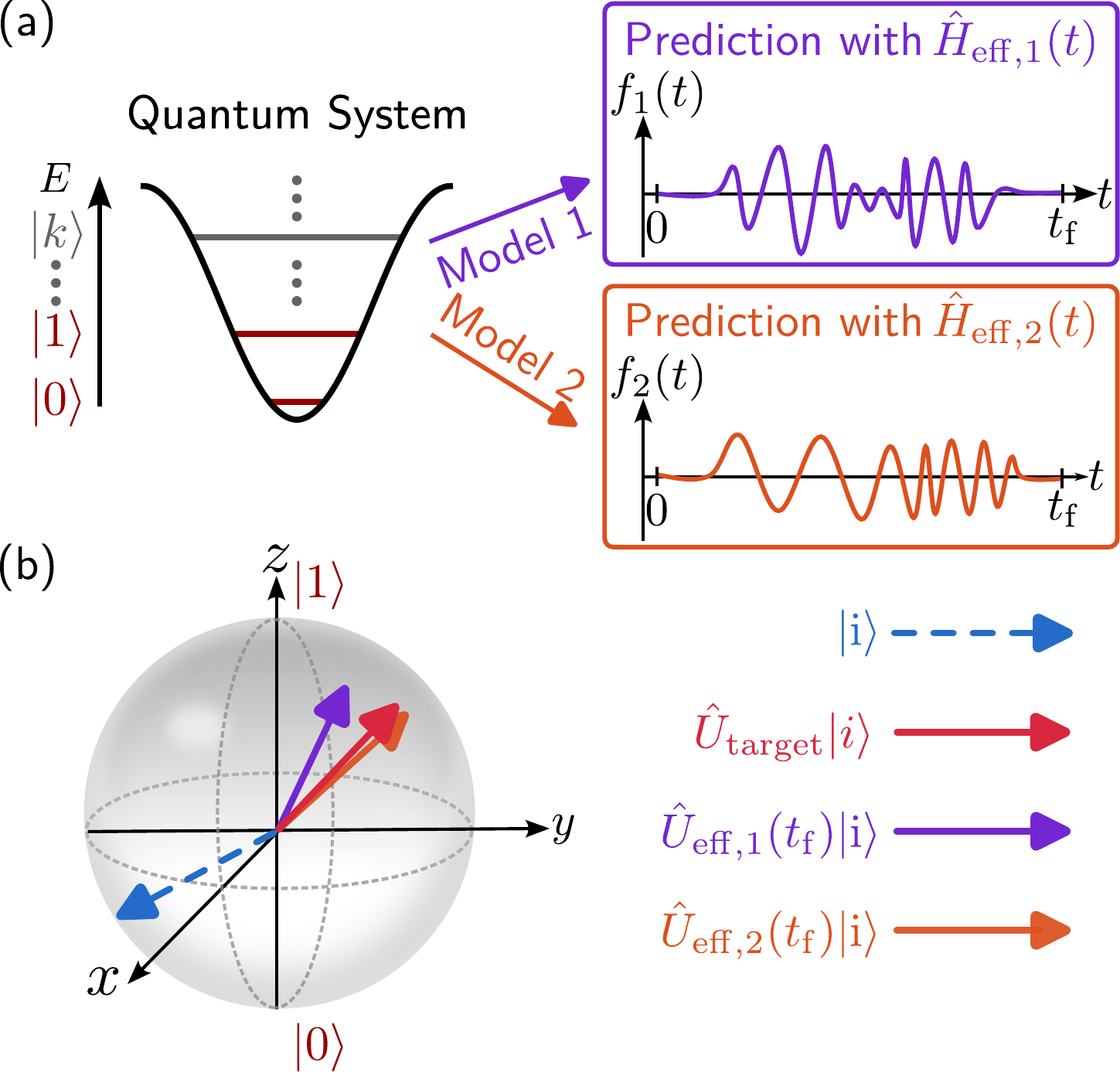}
    \caption{
        Model dependence of quantum-control predictions. (a) For a fixed target unitary $\hU_\mm{target}$, the same control
        framework can predict different control protocols when applied to different effective Hamiltonians. Two descriptions,
        $\hH_{\mm{eff},1}(t)$ and $\hH_{\mm{eff},2}(t)$, may therefore yield different controls, $f_1(t)$ and $f_2(t)$, designed
        to implement the same target operation. (b) When evaluated using a more faithful description of the physical system, the
        two protocols can generate different final unitaries and coherent errors. In the hybrid comparison considered below, the
        primitive pulse is calibrated independently in the model used for evaluation, while only the correction fields predicted
        by the approximate model are transferred.
    }
    \label{fig:CtrlModels}
\end{figure}

In this work, we investigate how the choice of effective Hamiltonian affects predictive pulse design for a capacitively driven
transmon. We compare two low-energy descriptions: a Hamiltonian obtained by diagonalizing the static transmon Hamiltonian and
expressing the physical charge drive in the resulting eigenbasis, and the standard Duffing approximation. We apply the same
correction-pulse construction to both models while calibrating the baseline pulse separately in each model to produce the same
nominal qubit rotation. The control method, target gate, and physical control channel are therefore held fixed, while the
effective Hamiltonian used to predict the multilevel correction is varied. This construction isolates model dependence in the
predicted correction from a trivial mismatch in the computational-subspace Rabi rate. We refer to the diagonalized-transmon
description as more faithful because it retains the low-energy spectrum and dressed charge matrix elements of the full cosine
Hamiltonian, while otherwise using the same rotating-wave treatment and Hilbert-space truncation procedure employed in the control
calculations.

Our first result is that correction fields calculated using the Duffing Hamiltonian can substantially reduce the gate error
predicted by that Hamiltonian but perform considerably worse when evaluated using the diagonalized-transmon description. To test
the transferability of these correction fields, we define a hybrid protocol in which the Duffing-derived correction quadratures
and constant detuning are combined with a baseline pulse independently calibrated using the diagonalized-transmon Hamiltonian.
Thus, the hybrid protocol transfers only the correction predicted by the Duffing model, not the Duffing-calibrated baseline pulse.
At short gate times, the hybrid protocol can fail to improve over the uncorrected diagonalized-transmon baseline and can even
increase the average fidelity error. This result shows that calibrating the nominal qubit rotation does not ensure that an
approximate Hamiltonian predicts the multilevel correction required by a more faithful description of the driven system.

We illustrate this dynamical model dependence by comparing the accumulated AC Stark phase predicted by the two Hamiltonians.
Individually small differences in the static spectrum and charge-drive matrix elements can combine over the duration of the gate
and produce different phase contributions to the dynamical error generator. The resulting AC Stark-phase mismatch provides a
transparent diagnostic of this effect, rather than a complete decomposition of the hybrid gate error. The mismatch increases as
$E_J/E_C$ is reduced, showing that the accuracy required of an effective Hamiltonian depends both on the device parameters and on
the driven control task for which the model is used.

Our second result is that the effective Hamiltonian also informs the choice of control framework. We first establish, using an
analytical sequential-transition estimate and numerical convergence tests, that a four-level diagonalized-transmon model captures
the relevant uncorrected and corrected dynamics over the gate-time range considered. Reducing the model to three levels, however,
removes higher-order error channels involving the second leakage level. Their omission can make a restricted correction strategy
appear sufficient, whereas the four-level model reveals relevant error channels that cannot be systematically suppressed within
that strategy at short gate times. We show that an extended construction addresses these additional channels using the same
physical control resources and produces a distinct control solution.

The paper is organized as follows. In Sec.~\ref{sec:EffectiveH}, we derive the diagonalized-transmon and Duffing Hamiltonians,
describe the model-specific calibration of the baseline pulse, and introduce a sequential-transition estimate for selecting the
Hilbert-space truncation. In Sec.~\ref{sec:OLQC}, we apply the same correction-pulse construction to both Hamiltonians and test
the transfer of Duffing-derived correction fields to an independently calibrated diagonalized-transmon baseline pulse. We also
compare the accumulated AC Stark phase predicted by the two Hamiltonians to illustrate how small model differences can produce
different dynamical error generators and correction fields. In Sec.~\ref{sec:ValidityTruncWithW}, we verify that the four-level
truncation remains valid after the correction fields are applied. In Sec.~\ref{sec:LinearCorrectionBreakdown}, we show that a
three-level truncation conceals error channels that limit the linear correction strategy in the four-level model. In
Sec.~\ref{sec:NonLinCorr}, we introduce the nonlinear correction strategy and show that it accesses a distinct control solution
using the same experimentally available controls.

\section{Low-energy effective Hamiltonians for a driven transmon qubit}
\label{sec:EffectiveH}

This section establishes three ingredients required for the control comparison:~The two effective Hamiltonians, the model-specific
calibration of the primitive gate, and the Hilbert-space truncation needed to capture the relevant leakage channels.

The Hamiltonian of a capacitively driven transmon is given by
\begin{equation}
    \hH (t) = 4 E_\uC \left(\hn + \frac{C_\ug [V_\mm{DC} + V (t)]}{2\ue}\right)^2 - E_\uJ \cos(\hat{\varphi}),
    \label{eq:HTransmon}
\end{equation}
where $E_\uC$ and $E_\uJ$ are, respectively, the charging and Josephson energies, and $2\ue$ is the charge of a Cooper pair. The
dynamical variables are the Cooper-pair number operator $\hn$ and the superconducting phase difference $\hat{\varphi}$ across the
Josephson junction. They obey $[\hat{\varphi},\hn]=i\mathbbm{1}$. Equation~\eqref{eq:HTransmon} is the standard circuit
Hamiltonian for the Cooper-pair-box family of superconducting qubits \cite{makhlin2001,koch2007,blais2021}. The transmon
regime corresponds to $E_\uJ/E_\uC\gg1$, where the charge dispersion of the low-lying levels is exponentially suppressed while
sufficient anharmonicity remains for selective qubit control~\cite{koch2007,schreier2008}.

We assume that the voltage source used to drive the transmon contains both a static component $V_\mm{DC}$ and a time-dependent
component $V (t)$. Our goal in this section is to construct low-energy effective Hamiltonians for this driven system and to
determine how many energy levels must be retained to describe the relevant dynamics quantitatively. We compare two descriptions.
The first is obtained by diagonalizing the static transmon Hamiltonian and then truncating the resulting eigenbasis. The second is
the Duffing approximation, obtained by expanding the cosine potential around one of its minima. Both descriptions lead to
infinite-dimensional Hamiltonians and therefore require a Hilbert-space truncation before they can be used in numerical
simulations or control calculations.

Throughout this work, we focus on coherent model mismatch arising from the Hamiltonian used to describe the driven transmon. We
take $V(t)$ to represent the time-dependent voltage waveform applied at the transmon control port and assume an ideal,
distortion-free mapping between the specified control envelope and this applied voltage. Decoherence, waveform distortion,
calibration drift, and control-line transfer-function uncertainty are not included. These effects constitute additional sources of
discrepancy between predicted and experimental dynamics and can, in principle, be incorporated into the control model or
robustness objective. Their omission here allows us to isolate the consequences of the effective Hamiltonian used to represent the
coherently driven system.

\subsection{Effective Hamiltonian via Truncation of the Transmon Eigenbasis}
\label{sec:HilbertSpaceTrunc}

We first split Eq.~\eqref{eq:HTransmon} into a static and a time-dependent drive Hamiltonian. We define
\begin{equation}
    \begin{aligned}
    n_g &=\frac{C_\ug V_\mm{DC}}{2\ue},\\
    \delta n_g(t) &=\frac{C_\ug V(t)}{2\ue}.
    \end{aligned}
\end{equation}
Expanding the quadratic term in Eq.~\eqref{eq:HTransmon} and dropping terms proportional to the identity, which only generate a
global phase, gives
\begin{equation}
    \begin{aligned}
        \hH (t) &= 4 E_\uC \left(\hn + n_g \right)^2 - E_J \cos(\hat{\varphi}) + f (t)  \hn,\\
        &= \hH_\uT + \hH_\ud (t)
    \end{aligned}
    \label{eq:HTransmonStDr}
\end{equation}
where $f(t)= 8 E_\uC \delta n_g(t)$. For a microwave drive near a carrier frequency $\omega_\ud$, we parametrize this
classical drive field in terms of two quadrature envelopes as
\begin{equation}
    f(t) = f_x (t) \cos(\omega_\ud t) + f_y (t) \sin(\omega_\ud t),
    \label{eq:DriveFct}
\end{equation}
where $f_x(t)$ and $f_y(t)$ are the in-phase and quadrature envelopes, respectively.

Let $\ket{k}$ denote the eigenstates associated to the eigenfrequencies $\omega_k$ of the static transmon Hamiltonian,
\begin{equation}
    \hH_\uT\ket{k}=\omega_k\ket{k},
\end{equation}
with $k \in \mathbbm{N}$. The transmon spectrum can be obtained either from the Mathieu-equation solution of the Cooper-pair-box
Hamiltonian or by numerical diagonalization in the charge basis~\cite{koch2007,schreier2008}. In this eigenbasis,
Eq.~\eqref{eq:HTransmonStDr} becomes
\begin{equation}
    \hH_\mm{diag} (t) = \sum_{k=0}^{\infty}\omega_k \ketbra{k}{k}
    + f(t) \sum_{k,l=0}^{\infty} n_{k,l}\ketbra{k}{l},
    \label{eq:HTransmonDiag}
\end{equation}
where $n_{k,l} = \bra{k} \hn \ket{l}$ are charge matrix elements. Thus, the capacitive drive couples to the charge operator
dressed by the exact transmon eigenbasis. Keeping this dressing is important for quantitative predictions of drive-induced
effects, such as leakage and AC Stark shifts in multilevel superconducting circuits~\cite{schuster2005,sank2016,dai2026}.

Before truncation and the rotating-frame approximations introduced below, Eq.~\eqref{eq:HTransmonDiag} is a change of basis of
Eq.~\eqref{eq:HTransmonStDr} and therefore describes the same coherent circuit Hamiltonian. Its advantage is that it makes the
low-energy structure and the corresponding dressed charge matrix elements explicit. However, the Hilbert space remains infinite
dimensional, whereas numerical simulations and pulse-design calculations require a finite-dimensional representation. For
transmons, the cutoff is commonly chosen by increasing the number of basis states or retained energy eigenstates until the
relevant spectral and dynamical quantities have converged~\cite{koch2007,groszkowski2021,larssen2026}. In driven control problems,
the relevant quantities include transition probabilities, leakage, and gate fidelities~\cite{motzoi2009,gambetta2011,hyyppa2024}.
Here, we complement this numerical convergence procedure with a leading-order Magnus estimate that identifies the dominant leakage
channels analytically and yields a physically transparent criterion for choosing the Hilbert-space cutoff.

\subsubsection{Rotating Frame and Interaction Picture}
\label{sec:IntPict}

We first transform Eq.~\eqref{eq:HTransmonDiag} to a frame rotating at the drive frequency. The unitary change-of-frame operator
is
\begin{equation}
    \hS_\mm{rot} (t) = \exp\left[-i \omega_\ud t \sum_{k=0}^{\infty} k \ketbra{k}{k} \right] =
    \sum_{k=0}^{\infty}e^{-i k\omega_\ud t} \ketbra{k}{k}.
    \label{eq:Urot}
\end{equation}
Using this transformation, the Hamiltonian in the rotating frame is
\begin{equation}
    \begin{aligned}
        \hH_\mm{rot} (t) &= \hS_\mm{rot}^\dagger(t)\hH_\mm{diag} (t) \hS_\mm{rot} (t) -i \hS_\mm{rot}^\dagger(t) \partial_t
        \hS_\mm{rot}(t) \\
        &= \sum_{k=0}^\infty \left[ \left(k -\frac{1}{2}\right) \Delta + \delta_k \right] \ketbra{k}{k} \\
        &\phantom{={}}
        + \frac{f_x (t)}{2} \sum_{m=0}^\infty \left(n_{m,m+1}\ketbra{m}{m+1} + H.c.\right) \\
        &\phantom{={}}
        + \frac{f_y (t)}{2} \sum_{m=0}^\infty \left(- i n_{m,m+1}\ketbra{m}{m+1} + H.c.\right),
    \end{aligned}
    \label{eq:Hrot}
\end{equation}
In Eq.~\eqref{eq:Hrot}, $\Delta=\omega_\uT-\omega_\ud$ is the detuning of the microwave drive from the
$\ket{0}\leftrightarrow\ket{1}$ transition, whose frequency $\omega_\uT=\omega_1-\omega_0$ is the transmon
frequency. The remaining diagonal terms encode the weak anharmonicity of the transmon spectrum. We write the remaining diagonal
terms in terms of the anharmonic level shift
\begin{equation}
	\delta_k=\omega_k-\omega_0-k\omega_\uT,
    \label{eq:CumulativeAnharmonicity}
\end{equation}
with $\delta_0 = \delta_1 =0$. The quantity $\delta_k$ measures the displacement of level $\ket{k}$ from a perfectly harmonic
spectrum with spacing $\omega_\uT$. It is also useful to define the anharmonicity of the $\ket{k-1}\leftrightarrow\ket{k}$
transition for $k\geq2$,
\begin{equation}
	\begin{aligned}
    	\alpha_k &= (\omega_k-\omega_{k-1})-\omega_\uT \\
        &= \delta_k-\delta_{k-1}.
	\end{aligned}
    \label{eq:TransitionAnharmonicity}
\end{equation}
The quantities $\delta_k$ determine the diagonal energies in the rotating frame. The detuning of the adjacent transition
$\ket{k}\leftrightarrow\ket{k+1}$ from the drive is $\Delta+\alpha_{k+1}$; on resonance with the qubit transition, this reduces to
$\alpha_{k+1}$.

In deriving Eq.~\eqref{eq:Hrot}, we have dropped a term proportional to the identity and performed the rotating-wave
approximation, neglecting counter-rotating terms oscillating at $2\omega_\ud$. The relevance of these terms can be assessed by
comparing the leading perturbative amplitudes of the corresponding error channels. For a gate of duration $t_\uf$, the leakage
amplitude associated with the weak anharmonicity scales as $1/(\abs{\alpha_2} t_\uf)$ whereas the amplitude generated by the rapidly
oscillating terms scales as $1/(\omega_\ud t_\uf)$. Their ratio obeys $\abs{\alpha_2}/\omega_\ud \ll 1$, since $\abs{\alpha_2}$ is one to
two orders of magnitude smaller than $\omega_\ud$ over the parameter range considered here. Errors generated by the rapidly
oscillating terms are parametrically smaller than the leakage errors retained in our effective Hamiltonian. At fixed dimensionless
gate time $\abs{\alpha_2} t_\uf$, the counter-rotating terms therefore do not contribute at the order relevant to the control errors
analyzed below. This perturbative hierarchy is consistent with the Magnus-based analysis of rapidly oscillating and leakage errors
developed in Ref.~\cite{roque2021}. We have also retained only the dominant charge matrix elements connecting adjacent transmon
eigenstates. At the charge-symmetry point, parity forbids charge matrix elements between eigenstates of equal parity. Among the
remaining symmetry-allowed matrix elements, the adjacent-level elements dominate because the low-energy transmon eigenstates
remain close to harmonic-oscillator states~\cite{koch2007,gambetta2011}. We have verified this hierarchy numerically for all
parameter sets considered below.

%We also retain only the dominant charge matrix elements connecting adjacent transmon eigenstates. At the charge-symmetry point,
%parity forbids charge matrix elements between eigenstates of equal parity. Among the remaining symmetry-allowed matrix elements,
%the adjacent-level elements dominate because the low-energy transmon eigenstates remain close to harmonic-oscillator states. We
%verify this hierarchy numerically for the parameter sets used below; representative values are provided in Table

We now decompose the rotating-frame Hamiltonian as $\hH_\mm{rot}(t)=\hH_0(t)+\hV(t)$. The Hamiltonian $\hH_0(t)$ generates the
desired dynamics in the computational subspace together with the free evolution of leakage states, while $\hV(t)$ describes the
remaining drive-induced couplings. For the truncation analysis below, we set $f_y(t)=0$, so that
\begin{equation}
    \hH_0 (t) = \sum_{k=0}^\infty \left[ \left(k-\frac{1}{2}\right)\Delta + \delta_k \right]\ketbra{k}{k} 
        + n_{0,1} \frac{f_x (t)}{2} \hsigma_x^{(0,1)},
    \label{eq:Hrot0}
\end{equation}
and
\begin{equation}
    \hV (t) = \frac{f_x (t)}{2} \sum_{m=1} n_{m,m+1} \hsigma_x^{(m,m+1)}.
    \label{eq:Vrot}
\end{equation}
Here, we have used $n_{k,l}=n_{l,k}$ and introduced the generalized Pauli operators
\begin{equation}
    \begin{aligned}
        \hsigma_x^{(m,n)}
        &=
        \ketbra{m}{n}+\ketbra{n}{m},
        \\
        \hsigma_y^{(m,n)}
        &=
        % this is the std def
        i\ketbra{m}{n} - i\ketbra{n}{m},
        % what I had
        %i\ketbra{m}{n} - i\ketbra{n}{m},
        \\
        \hsigma_z^{(m,n)}
        &=
        \ketbra{n}{n}-\ketbra{m}{m},
    \end{aligned}
    \label{eq:GenPauli}
\end{equation}
with the convention $m < n$. Using these generalized Pauli operators, the $f_y (t)$ term in Eq.~\eqref{eq:Hrot} is proportional to
$\hsigma_y^{(m,m+1)}$.

When the drive is resonant with the qubit transition, $\Delta=0$, the Hamiltonian $\hH_0(t)$ commutes with itself at different
times. The corresponding unitary evolution is thus
\begin{equation}
    \begin{aligned}
        \hU_0 (t) &= \exp\left[ -i \int_0^t \di{t_1} \hH_0 (t_1) \right] \\
        &=\cos\left[ \frac{\theta (t)}{2 } \right] \mathbbm{1}_\mm{comp} - i \sin\left[ \frac{\theta (t)}{2}\right]\hsigma_x^{(0,1)}\\
        &\phantom{={}}
        + \sum_{k \geq 2} \exp(-i \delta_k t) \ketbra{k}{k},
    \end{aligned}
    \label{eq:U0rot}
\end{equation}
where $\theta(t) = n_{0,1} \int_0^t \di{t_1} f_x(t_1)$ is the rotation angle of the $x$ rotation in the computational subspace,
and $\mathbbm{1}_\mm{comp}=\ketbra{0}{0}+\ketbra{1}{1}$ is the computational-subspace identity. Choosing the pulse area such that
$\theta(t_\uf)=\pi/2$ leads to $\hU_0 (t_\uf)$ being the complex Hadamard gate used below.

To estimate leakage, we now move to the interaction picture generated by $\hU_0(t)$. This transformation removes the intended
qubit rotation and the free evolution of the leakage states. The remaining interaction-picture Hamiltonian $\hV_\uI (t) =
\hU_0^\dagger(t)\hV(t)\hU_0(t)$ isolates the residual couplings that drive population out of the computational subspace. Using
Eqs.~\eqref{eq:Vrot} and \eqref{eq:U0rot}, we obtain
\begin{equation}
    \begin{aligned}
        \hV_\uI (t) &= \frac{f_x(t) }{2} \left\{ \sum_{k=2}^\infty n_{k,k+1} \left[\cos\left( \alpha_{k+1} t \right)
            \hsigma_x^{(k,k+1)} \right. \right. \\
            &\phantom{={}}
            - \left. \sin\left( \alpha_{k+1} t \right) \hsigma_y^{(k,k+1)} \right]  \\
            &\phantom{={}}
            + n_{1,2}  \left[ \cos\left[ \theta/2 (t) \right] \left( \cos\left( \delta_2 t \right)
            \hsigma_x^{(1,2)} - \sin\left( \delta_2 t \right) \hsigma_y^{(1,2)} \right) \right. \\
            &\phantom{={}} 
            \left. \vphantom{\sum_{k=2}} 
            \left. + \sin\left[ \theta/2 (t) \right] \left( \sin\left( \delta_2 t \right)  \hsigma_x^{(0,2)} + \cos\left(
            \delta_2 t \right) \hsigma_y^{(0,2)} \right) \right] \right\},
    \end{aligned}
    \label{eq:HrotIntPict}
\end{equation}
Equation~\eqref{eq:HrotIntPict} shows how the drive couples neighboring states in the transmon energy ladder once the intended
qubit dynamics has been removed. The terms proportional to $n_{1,2}$ couple the computational states to the first leakage level
$\ket{2}$, while the terms proportional to $n_{k,k+1}$ with $k\geq 2$ couple neighboring leakage levels. Although only neighboring
levels are coupled directly in $\hV_\uI(t)$, the resulting time evolution contains higher-order leakage pathways. For example, the
sequence $\ket{0}\to \ket{2}\to \ket{3}$ produces a second-order contribution to the effective transition amplitude $\ket{0}\to
\ket{3}$. The Magnus expansion provides a systematic way to identify and rank these higher-order leakage channels, which is the
basis for the Hilbert-space truncation criterion derived below.

In the following, we use the pulse envelope
\begin{equation}
    f_x (t) = \frac{\theta_0}{n_{0,1} t_\uf}\left[1 - \cos \left( 2\pi \frac{t}{t_\uf} \right)\right],
    \label{eq:fx}
\end{equation}
where $\theta_0$ is the desired rotation angle. The factor $n_{0,1}$ in Eq.~\eqref{eq:fx} models the calibration of the nominal
qubit rotation. Experimentally, the charge-operator matrix element $n_{0,1}$ need not be determined independently because the
applied microwave amplitude is calibrated to produce a specified rotation angle. We implement this calibration separately within
each effective model by evaluating $n_{0,1}$ using that model. Consequently, the Duffing and diagonalized-transmon baseline pulses
are normalized using their respective charge matrix elements, and both produce the same nominal rotation angle because the
computational-subspace coupling is proportional to $n_{0,1} f_x (t)$. 

This model-specific calibration removes the elementary discrepancy in the nominal qubit Rabi rate. It does not calibrate the
multilevel driven response, which depends on the relative higher-level charge matrix elements and transition detunings. These
model-dependent quantities jointly determine the dynamical error generator and, consequently, the correction fields predicted by
the control construction.

\subsubsection{Magnus Expansion and Transition Probabilities to Leakage States}
\label{sec:AnalyticsTrunc}

We now use Eq.~\eqref{eq:HrotIntPict} to estimate which leakage levels must be retained in the effective Hilbert space. The
time-evolution operator generated by $\hV_\uI(t)$ can be written as
\begin{equation}
    \hU_\uI(t)=\exp[\hOmega(t)],
    \label{eq:UIMagnus}
\end{equation}
where $\hOmega(t)=\sum_{l=1}^{\infty}\hOmega_l(t)$ is the Magnus expansion~\cite{magnus1954,blanes2009}. The first two terms are
\begin{equation}
    \begin{aligned}
        \hOmega_1(t) &= -i\int_0^t \di{t_1}\,\hV_\uI(t_1), \\
        \hOmega_2(t) &= \frac{1}{2} \int_0^t \di{t_1} \left[\partial_{t_1}\hOmega_1(t_1), \hOmega_1(t_1) \right].
    \end{aligned}
    \label{eq:MagnusExp2ndOrd}
\end{equation}
This representation organizes leakage processes according to the number of elementary drive-induced couplings involved. Matrix
elements of $\hOmega_1(t_\uf)$ describe processes generated by a single application of the interaction Hamiltonian, while
higher-order Magnus terms contain processes built from multiple such couplings.

Equation~\eqref{eq:HrotIntPict} directly couples the computational subspace to the first leakage level $\ket{2}$ and couples
neighboring leakage levels to one another. Population transfer to higher levels therefore occurs through sequences of adjacent
transitions. For example, the pathway $\ket{0} \to \ket{2} \to \ket{3}$ motivates the sequential-transition estimate
\begin{equation}
    \eta_{0 \to 3} \sim P^{(1)}_{\ket{0}\to\ket{2}} P^{(1)}_{\ket{2}\to\ket{3}},
    \label{eq:EstTransProb03}
\end{equation}
where
\begin{equation}
    P^{(1)}_{\ket{i}\rightarrow\ket{j}} = \abs{\bra{j}\hOmega_1(t_\uf)\ket{i}}^2.
    \label{eq:1stOrdMagnusProb}
\end{equation}
is the transition probability obtained from the first-order Magnus expansion.

Generalizing this construction to successive adjacent transitions gives the conservative sequential-transition estimate
\begin{equation}
    \begin{aligned}
        \eta_k &= 16 \frac{n_{1,2}^2}{n_{0,1}^2} \frac{\pi^4 \theta_0^2}{\abs{\alpha_2 t_\uf}^6} \\
        &\phantom{={}} 
        \times \prod_{m=3}^{k} 16 \frac{n_{m-1,m}^2}{n_{0,1}^2}
        \frac{\pi^4 \theta_0^2}{\left(\alpha_{m} t_\uf\right)^2 \left[-4\pi^2 + \left(\alpha_{m} t_\uf\right)^2\right]^2}.
    \end{aligned}
    \label{eq:CutOffCond}
\end{equation}
Here, we use the pathway originating from $\ket{1}$ because, as shown in Appendix~\ref{app:Omega1MatrixElements}, its upper bound
for the initial transition into $\ket{2}$ is four times larger than the corresponding bound for a pathway originating from
$\ket{0}$. This choice therefore gives the more conservative sequential-transition estimate.

Because $\eta_k$ is constructed from products of adjacent first-order Magnus transition probabilities, it is intended to rank
successive leakage channels rather than to approximate an exact higher-order transition probability. A truncation retaining states
through $\ket{k_\mm{cut}}$ is expected to be sufficient when the sequential-transition estimate for the first omitted level is
parametrically smaller than that for the highest retained level, $\eta_{k_\mm{cut}+1} \ll \eta_{k_\mm{cut}}$. The estimate
contains the physical ingredients controlling the leakage hierarchy:~The dressed charge matrix elements, transition
anharmonicities, gate duration, and pulse smoothness.

For the diagonalized-transmon Hamiltonian, we evaluate this criterion at the shortest gate time considered, $|\alpha_2|t_f=5.74$.
We find $\eta_4/\eta_3=2.99\times10^{-4}$, $3.26\times10^{-4}$, and $1.10\times10^{-3}$ for $E_J/E_C=50$, $40$, and $30$,
respectively. Thus, for all three parameter sets, the sequential-transition estimate for the first omitted level is suppressed by
approximately three or more orders of magnitude relative to that for the highest retained level. This supports a four-level
truncation over the parameter range considered here.

\begin{figure}[t]
    \includegraphics[width=0.99\columnwidth]{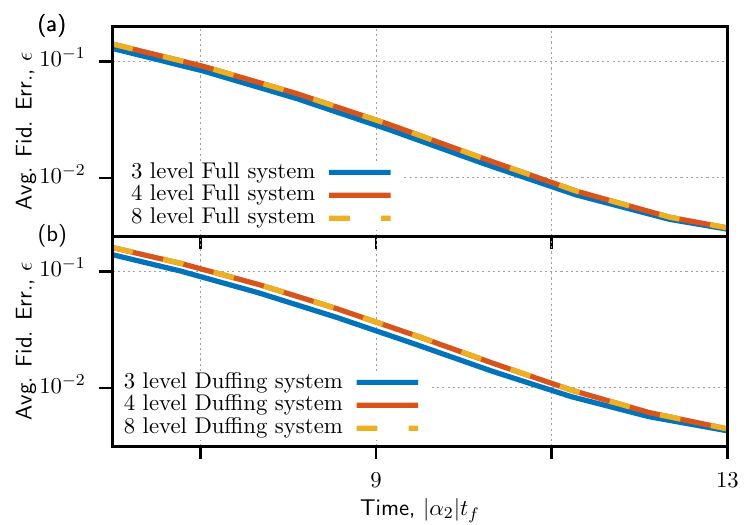}
    \caption{
        Hilbert-space convergence of the uncorrected single-qubit gate. Average fidelity error $\epsilon$ for the target
        complex-Hadamard gate as a function of the dimensionless gate time $\abs{\alpha_2}t_\uf$. (a) Results obtained from the
        diagonalized-transmon Hamiltonian truncated to three, four, and eight energy levels. (b) Corresponding results for the
        Duffing Hamiltonian truncated to the same number of levels. In both models, the four- and eight-level calculations are nearly
        indistinguishable over the gate-time range shown, while the three-level truncation gives visible quantitative deviations.
        This establishes the four-level truncation as sufficient for the control comparisons considered below.
    }
    \label{fig:FidelityOfDifferentDimensionality}
\end{figure}

To validate our approach, we plot in Fig.~\ref{fig:FidelityOfDifferentDimensionality}~(a) as a function of dimensionless time
$\abs{\alpha_2} t_\uf$ the average fidelity error~\cite{pedersen2007} 
\begin{equation}
    \epsilon = 1- F=\frac{1}{d(d+1)}\left[\mm{Tr}(\hO \hO^\dagger) + \abs{\mm{Tr}(\hO)}^2\right],
\label{eq:fid}
\end{equation}
when the desired gate is a complex Hadamard gate. In Eq.~\eqref{eq:fid}, $\hO=\hP \hU_\mm{Ideal}^\dagger \hU(t_\uf) \hP $ serves
as an operator that compares the target unitary 
\begin{equation}
    \hU_\mm{Ideal} = \frac{1}{\sqrt{2}}\left( \ketbra{0}{0} -i \ketbra{0}{1} - i \ketbra{1}{0} + \ketbra{1}{1} \right)
    \label{eq:UId}
\end{equation}
with the evolution $\hU (t_\uf)$ generated by the system Hamiltonian, after both are restricted to the computational subspace
through the projection operator $\hP$. We compare the average fidelity error obtained by truncating the Hilbert space associated
to Eq.~\eqref{eq:Hrot} to three (solid blue line), four (solid orange line), and eight energy levels (dashed yellow line). Our
results show that there is no significant quantitative difference between the results obtained with four or eight energy levels,
further validating the proposed analytical approach for truncating the Hilbert space.

\subsection{Duffing Approximation}
\label{sec:DuffApprox}

We now compare the diagonalized-transmon description of Sec.~\ref{sec:HilbertSpaceTrunc} with the commonly used Duffing
approximation. The Duffing model is widely used to describe transmon qubits as weakly anharmonic oscillators, and it provides a
convenient starting point for analytical pulse-design methods in superconducting
circuits~\cite{koch2007,motzoi2009,gambetta2011,blais2021,hyyppa2024}. The purpose of the comparison here is not only to introduce
a simpler low-energy model, but also to identify which physical features are modified by this approximation. As we show below, the
Duffing approximation modifies both the static transmon spectrum and the matrix elements of the physical charge-drive operator.
These two changes are central to the model dependence of the correction pulses studied in Sec.~\ref{sec:OLQC}.

The Duffing approximation is obtained by expanding the Josephson potential around one of its minima. This procedure is controlled
in the transmon regime, $E_J/E_C \gg 1$, where the low-lying wave functions are localized near the bottom of a cosine
well~\cite{koch2007,schreier2008,blais2021}. In this regime, the offset charge $n_g$ can be removed from the local Hamiltonian by
the gauge transformation $\Psi(\varphi)=\exp[-i n_g\varphi]\psi(\varphi)$. The offset charge then enters only through the boundary
conditions, whose effect on the low-energy spectrum is exponentially suppressed.  We therefore neglect charge dispersion in the
Duffing model, consistent with the standard transmon approximation~\cite{koch2007,schreier2008}.

Starting from Eq.~\eqref{eq:HTransmonStDr}, we expand the Josephson potential and truncate the result at quartic order, dropping
terms proportional to the identity:
\begin{equation}
    \begin{aligned}
        \hH_\mm{Duff} (t) &= 4 E_\uC \hn^2 - E_\uJ \sum_{k=0}^{\infty} \frac{(-1)^k}{(2k)!}\hat{\varphi}^{2k} + f (t) \hn,\\
        &= 4 E_\uC \hn^2 + \frac{E_J}{2}\hat{\varphi}^2 - \frac{E_J}{24}\hat{\varphi}^4 + f (t) \hn
        +\mathcal{O}\left(\hat{\varphi}^6\right).
    \end{aligned}
    \label{eq:Hflux2}
\end{equation}
This is the usual quartic approximation to the Cooper-pair-box Hamiltonian in the large $E_J/E_C$ regime
\cite{koch2007,makhlin2001,blais2021}, where terms proportional to identity have been dropped.

We introduce harmonic-oscillator operators according to
\begin{equation}
    \begin{aligned}
        %\hat{n} &= -\frac{i}{2} \left(\frac{E_\uJ}{2 E_\uC}\right)^\frac{1}{4} \left(\hb-\hb^\dagger\right), \\
        %\hat{\varphi} &= \left(\frac{2 E_\uC}{E_\uJ}\right)^\frac{1}{4} \left(\hb+\hb^\dagger\right),
        \hat{n} &= i n_\mm{zpf} \left(\hb^\dagger - \hb \right),\\
        \hat{\varphi} &= \varphi_\mm{zpf} \left( \hb^\dagger + \hb \right),
    \end{aligned}
    \label{eq:2ndQuant}
\end{equation}
with 
\begin{equation}
    \begin{aligned}
        n_\mm{zpf} &= \frac{1}{2} \left(\frac{E_\uJ}{2 E_\uC}\right)^\frac{1}{4} \\
        \varphi_\mm{zpf} &= \left(\frac{2 E_\uC}{E_\uJ}\right)^\frac{1}{4}.
    \end{aligned}
    \label{eq:DuffingZPF}
\end{equation}
The quantities $\varphi_\mm{zpf}$ and $n_\mm{zpf}$ are the zero-point fluctuation amplitudes of the phase and charge variables in
the harmonic approximation to the transmon potential. They set the natural oscillator length scales through $\langle
0|\hat{\varphi}^2|0\rangle=\varphi_\mm{zpf}^2$ and $\langle 0|\hat n^2|0\rangle=n_\mm{zpf}^2$. In the transmon regime $E_J/E_C \gg
1$, one has $\varphi_\mm{zpf} \ll 1$, so the phase wave function is localized near a minimum of the cosine potential, while the
conjugate charge exhibits correspondingly larger zero-point fluctuations. Thus, the smallness of $\varphi_\mm{zpf}$ provides the
expansion parameter controlling the quartic approximation to the Josephson potential, whereas $n_\mm{zpf}$ fixes the scale of the
charge-drive matrix elements in the Duffing approximation.

Substituting Eq.~\eqref{eq:2ndQuant} into Eq.~\eqref{eq:Hflux2} and retaining only the number-conserving part of the static
Hamiltonian gives the Duffing Hamiltonian
\begin{equation}
    \hH_\mm{Duff} (t) = \omega_\uT \hb^\dagger \hb  - \frac{\omega_\uC}{2}
    \hb^\dagger \hb \left( \hb^\dagger \hb - \mathbbm{1} \right) + i n_\mm{zpf} f(t) \left(\hb^\dagger - \hb\right).
    \label{eq:Duffing1}
\end{equation}
Here, we have written the Hamiltonian in angular-frequency units, with $\omega_\uC=E_\uC/\hbar$, and have used the same symbol
$f(t)$ for the drive amplitude after division by $\hbar$. The frequency $\omega_\uT = \omega_\up -\omega_\uC = (\sqrt{8E_\uC
E_\uJ}-E_\uC)/\hbar$ is the Duffing approximation to the transmon qubit frequency, while $\omega_\up = \sqrt{8E_\uC
E_\uJ}/\hbar$ is the plasma frequency associated with small oscillations about a minimum of the transmon potential. Physically,
the quartic term both shifts the harmonic frequency from $\omega_\up$ to $\omega_\uT=\omega_\up-\omega_\uC$ and generates the Kerr
nonlinearity.

It follows directly from Eq.~\eqref{eq:Duffing1} that the transition frequency between level $\ket{k+1}$ and $\ket{k}$ is
$\omega_\uT- k \omega_\uC$. Thus, the Duffing model imposes a constant adjacent-transition anharmonicity $\alpha = -\omega_\uC$, in
contrast to the diagonalized-transmon model of Sec.~\ref{sec:HilbertSpaceTrunc}, where the transition frequencies are obtained
from the full cosine Hamiltonian rather than imposed by the quartic expansion~\cite{koch2007,schreier2008,groszkowski2021}.

We can now transform Eq.~\eqref{eq:Duffing1} to the frame rotating at the drive frequency. After applying the rotating-wave
approximation with respect to the drive frequency, we obtain
\begin{equation}
    \begin{aligned}
        \hH_\mm{Duff,rot} (t) &= \Delta \hb^\dagger\hb - \frac{\omega_\uC}{2} \hb^\dagger \hb \left( \hb^\dagger
        \hb - \mathbbm{1} \right) \\
        %&\phantom{={}} + \frac{i}{2}\left(\frac{\omega_\uJ}{2 \omega_\uC}\right)^\frac{1}{4} \frac{f_x (t)}{2}\left(\hb^\dagger - \hb\right),
        &\phantom{={}} + \frac{i}{2} n_\mm{zpf} \left[f_x (t) \left(\hb^\dagger - \hb\right) + i f_y (t) \left(\hb^\dagger +
        \hb\right)\right],
    \end{aligned}
\label{eq:Duffing2}
\end{equation}
where $\Delta = \omega_\uT - \omega_\ud$ is the detuning between the transmon and drive frequencies.

\begin{figure}[t]
    \includegraphics[width=0.99\columnwidth]{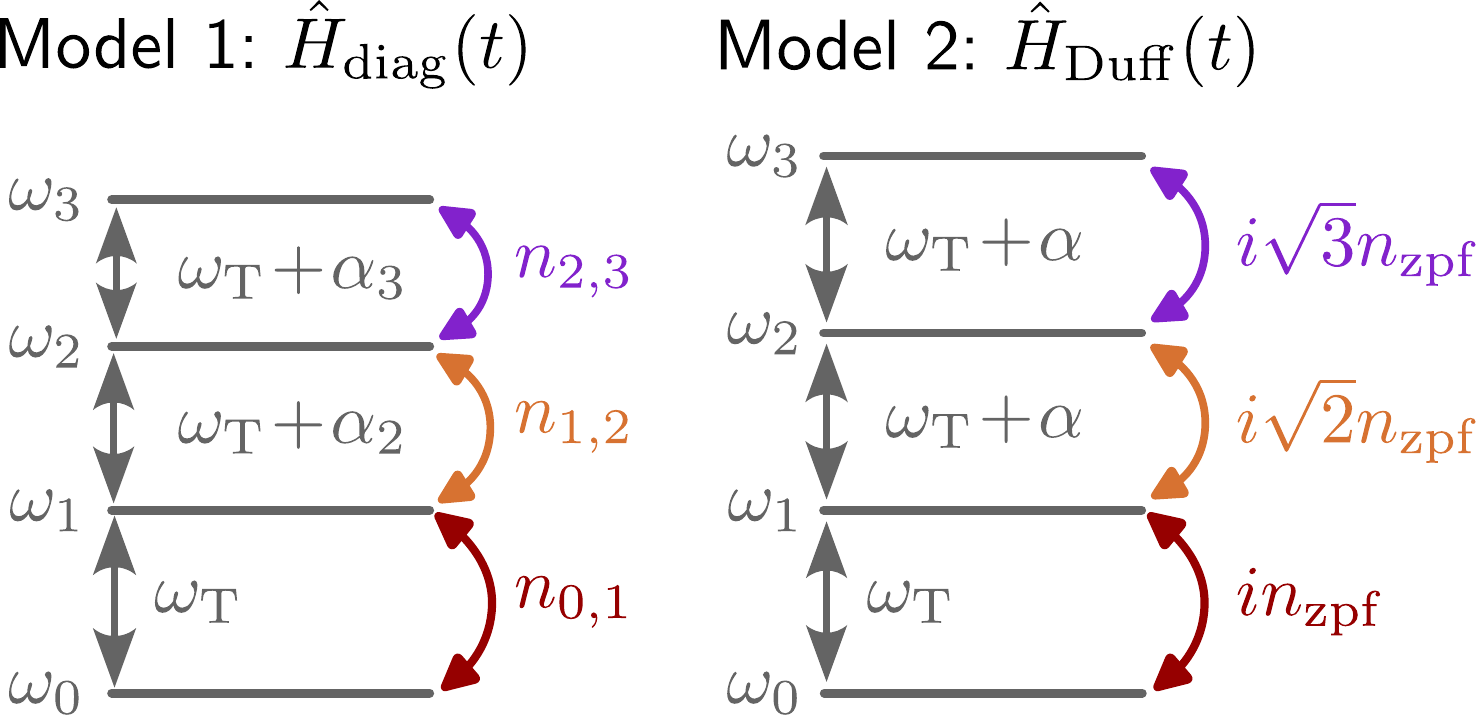}
    \caption{
    Comparison of the two effective transmon models used for pulse design. The diagonalized-transmon model, $\hH_\mm{diag} (t)$,
    retains the transition frequencies and dressed charge matrix elements $n_{k,l}$ obtained from the eigenbasis of the static transmon
    Hamiltonian. The Duffing model, $\hH_\mm{Duff} (t)$, replaces the spectrum by that of a Kerr oscillator with constant
    adjacent-transition anharmonicity and represents the charge-drive matrix elements by their harmonic-oscillator values,
    proportional to $\sqrt{k+1} n_\mm{zpf}$. Thus, the two models differ both in their level spacings and in the control matrix
    elements through which the microwave drive couples adjacent states.
    }
    \label{fig:Qcircuit}
\end{figure}

Equation~\eqref{eq:Duffing2} makes explicit the two approximations that distinguish the Duffing model from the
diagonalized-transmon Hamiltonian [see Fig.~\ref{fig:Qcircuit}]. First, the static spectrum is replaced by that of a Kerr
oscillator with constant anharmonicity between neighboring energy states. Second, the physical voltage drive is represented by the
harmonic-oscillator charge operator. In the Duffing approximation, the magnitude of the adjacent-level charge matrix element is
therefore fixed by
\begin{equation}
    \bra{k+1} \hat{n} \ket{k} = i \sqrt{k+1} n_\mm{zpf},
    \label{eq:duff_drive_matrix_elements}
\end{equation}
up to the phase convention used for the oscillator states. By contrast, in the diagonalized-transmon model the same physical drive
couples through the dressed charge matrix elements $n_{k,l}$ [see Eq.~\eqref{eq:HTransmonDiag}]. Consequently, the Duffing and
diagonalized-transmon Hamiltonians differ not only in their transition frequencies, but also in the representation of the control
operator in the truncated Hilbert space. Normalizing the baseline pulse by the model-specific value of $n_{0,1}$ removes the
direct discrepancy in the nominal qubit Rabi rate. The remaining model dependence is associated with the multilevel driven
dynamics, including the relative higher-level charge matrix elements, the transition detunings, and the way these quantities enter
the full time-dependent error generator. In the Duffing model, the higher-level matrix elements follow their harmonic-oscillator
values, whereas in the diagonalized-transmon model they are dressed by the full cosine potential. Consequently, a correction field
calculated from the Duffing dynamics need not produce the same correction when coupled through the charge operator of the
diagonalized-transmon model. This distinction is particularly relevant for fast gates, where leakage, drive-induced phase shifts,
and unwanted transitions are enhanced by stronger microwave drives~\cite{motzoi2009,gambetta2011,hyyppa2024,dai2026}.

These model-dependent differences are important for pulse design. Correction fields computed using the Duffing Hamiltonian are
constructed to cancel the error generator associated with Eq.~\eqref{eq:Duffing2}. However, the physical transmon dynamics is more
accurately represented by the diagonalized Hamiltonian, where both the level spacings and the charge-drive matrix elements are
dressed by the full cosine potential. Correction fields that are self-consistent for the Duffing error generator therefore need
not cancel the error generator associated with the diagonalized-transmon model. The comparison in Sec.~\ref{sec:OLQC} tests
precisely this point by constructing correction fields from the Duffing Hamiltonian and then adding those fields to a
diagonalized-transmon baseline pulse that has been independently calibrated using $n_{0,1}$ obtained from the
diagonalized-transmon Hamiltonian. The model-dependent AC Stark shifts discussed later provide one concrete manifestation of this
mismatch, consistent with the known sensitivity of driven superconducting circuits to drive-induced frequency shifts and
noncomputational transitions~\cite{schuster2005,sank2016,dai2026}.

Like the diagonalized-transmon Hamiltonian, the Duffing Hamiltonian acts on an infinite-dimensional Hilbert space and must be
truncated for numerical simulations. Applying the same sequential-transition estimate used in Sec.~\ref{sec:HilbertSpaceTrunc}
shows that retaining states through $\ket{3}$, i.e., using a four-level model, is sufficient for the gate times considered here.
For the Duffing Hamiltonian, the ratio $\eta_4/\eta_3$ is independent of $E_J/E_C$ when evaluated at fixed dimensionless gate time
$\abs{\alpha_2} t_\uf$. This follows from the structure of the Duffing model:~The adjacent charge matrix elements have the
harmonic-oscillator form $n_{k,k+1}\propto\sqrt{k+1}\,n_\mm{zpf}$, so the relevant ratios of matrix elements are independent of
$E_J/E_C$, while the transition anharmonicities are fixed by the constant Duffing anharmonicity. Thus, the dependence on the
absolute scale of the anharmonicity cancels in Eq.~\eqref{eq:CutOffCond} when the criterion is expressed in terms of the
dimensionless quantity $\abs{\alpha_2} t_\uf$. At $\abs{\alpha_2} t_\uf=5.74$, we find $\eta_4/\eta_3=1.99\times10^{-3}$ for the
Duffing model, so the sequential-transition estimate for the first omitted level remains parametrically smaller than that for the
highest retained level. This again supports the use of a four-level truncation.

We confirm this numerically by computing the average fidelity error as a function of dimensionless time $\abs{\alpha_2} t_\uf$
when the target unitary is given by Eq.~\eqref{eq:UId}. This corresponds to setting $f_y (t) =0$ in Eq.~\eqref{eq:Duffing2} and
using $f_x (t)$ as given in Eq.~\eqref{eq:fx}. The results are shown in Fig.~\ref{fig:FidelityOfDifferentDimensionality}~(b). The
four-level (orange trace) and eight-level (dashed yellow trace) give nearly indistinguishable average fidelity errors for the
range shown, whereas the three-level truncation (blue trace) shows visible quantitative deviations. We therefore use a four-level
Duffing model in the control comparisons below.

\section{Impact of Effective Hamiltonian on Open-Loop Quantum Control}
\label{sec:OLQC}

The analysis above provides two effective Hamiltonians for the driven low-energy dynamics of a transmon qubit. In both
descriptions, the infinite-dimensional Hilbert space must be truncated before performing numerical simulations or designing
control pulses. The truncation criterion discussed in Secs.~\ref{sec:HilbertSpaceTrunc} and \ref{sec:DuffApprox} shows that, for
the gate times considered below, a four-level description is sufficient to capture the relevant leakage dynamics [see
Fig.~\ref{fig:FidelityOfDifferentDimensionality}]. Having fixed the effective Hilbert space, we now
turn to the construction of correction pulses.  Rather than developing a new control protocol, we use the Magnus-based
constrained-control framework introduced in Refs.~\cite{ribeiro2017,roque2021} and apply it separately to the two effective
Hamiltonians derived above. This allows us to isolate how the choice of effective Hamiltonian affects the pulse predicted by an
otherwise fixed control strategy.

The Magnus-based control framework~\cite{ribeiro2017,roque2021} starts from a decomposition of the Hamiltonian into an ideal
contribution $\hH_0(t)$, which generates the target gate, and residual terms $\hV(t)$, which generate unwanted dynamics such as
leakage and phase errors. 

We restrict the correction protocol to experimentally available control parameters. For the capacitively driven transmon
considered here, these are the two microwave quadratures and the drive frequency. The latter does not appear as an additional term
in the lab-frame correction Hamiltonian, but it translates, after transforming to the rotating frame, into the detuning $\Delta$
appearing in Eq.~\eqref{eq:Hrot}. In the uncorrected protocol, the drive is resonant with the qubit transition, corresponding to
$\Delta=0$. In the corrected protocol, however, we allow the drive frequency $\omega_\ud$ to be chosen such that the
rotating-frame dynamics has a nonzero detuning. Following the constrained-control strategy of Ref.~\cite{roque2021}, we restrict
this detuning to be constant during the gate, rather than allowing a time-dependent chirp of the drive frequency, since a fixed
detuning is simpler to implement and calibrate experimentally. The modified Hamiltonian is written as
\begin{equation}
    \hH_\mm{mod}(t)=\hH_0(t)+\hV(t)+\hW(t),
    \label{eq:Hmod}
\end{equation}
where the lab-frame correction Hamiltonian is applied through the charge drive,
\begin{equation}
  \hW(t) = \left[ g_x(t) \cos(\omega_\ud t) + g_y(t) \sin(\omega_\ud t) \right]\hn ,
  \label{eq:W}
\end{equation}
Here, $g_x (t)$ and $g_y (t)$ are correction quadrature envelopes, while the choice of $\omega_\ud$ fixes the constant detuning
$\Delta$ in the rotating-frame description. Equivalently, $\hH_\mm{mod}(t)$ has the same form as the original driven Hamiltonian,
but with the drive envelopes replaced by
\begin{equation}
    \begin{aligned}
        f_x(t) &\to f_{\mm{mod},x}(t) = f_x(t) + g_x(t), \\
        f_y(t) &\to f_{\mm{mod},y}(t) = f_y(t) + g_y(t),
    \end{aligned}
    \label{eq:ModEnv}
\end{equation}
see Eq.~\eqref{eq:DriveFct}. In our implementation, the envelopes $g_x(t)$ and $g_y(t)$ are expanded in a finite Fourier basis.
This parametrization provides enough freedom to shape the correction pulse while preserving the experimental constraint that no
additional control operators are introduced.

The pulse coefficients are determined in the interaction picture defined by the ideal propagator $\hU_0(t)$ generated by
$\hH_0(t)$. In this frame, the desired gate is obtained when the residual propagator associated with the modified Hamiltonian
satisfies
\begin{equation}
    \hU_\mm{mod,I}(t_f) = \hT \exp \left[-i\int_0^{t_f}\di{t} \hH_\mm{mod,I}(t) \right] = \mathbbm{1},
    \label{eq:UICond}
\end{equation}
so that the full evolution at the final time coincides with the target operation. Writing the interaction-picture propagator as a
Magnus expansion,
\begin{equation}
    \hU_\mm{mod,I}(t_f) = \exp\left[ \sum_{l=1}^\infty \hOmega_l(t_\uf) \right],
    \label{eq:Magnus}
\end{equation}
the correction conditions are imposed perturbatively by requiring
\begin{equation}
    \sum_{l=1}^m \hOmega_l (t_f) = \mathbf{0}
    \label{eq:MagnusCond}
\end{equation}
up to a chosen order $m$. This yields a set of algebraic equations for the Fourier coefficients defining the correction pulse.
The perturbative construction and its domain of applicability were established in Refs.~\cite{ribeiro2017,roque2021}. In the
present work, the resulting correction pulses are evaluated by direct numerical propagation of the corresponding system
Hamiltonian, providing an operational validation of the construction over the parameter range considered below.

In this work, the role of the Magnus framework is not to introduce a new control algorithm, but to compare the control pulses
predicted by different effective Hamiltonians. We use the truncated Hamiltonian $\hH_\mm{diag}(t)$ obtained from diagonalizing the
static transmon Hamiltonian as the more accurate low-energy model of the physical system. This makes it possible to test the
Duffing approximation directly. We first compute the correction fields using $\hH_\mm{Duff}(t)$. We then combine those
Duffing-derived correction fields with a baseline pulse independently calibrated using the diagonalized-transmon value of
$n_{0,1}$, and evaluate the resulting hybrid protocol using $\hH_\mm{diag}(t)$. As shown below, this procedure yields
substantially poorer gate fidelities, demonstrating that correction fields that appear successful within the Duffing model need
not remain effective when added to a calibrated baseline pulse in the diagonalized-transmon model.

\subsection{Model dependence of Magnus-based correction pulses}

\begin{figure}[t]
    \includegraphics[width=0.99\columnwidth]{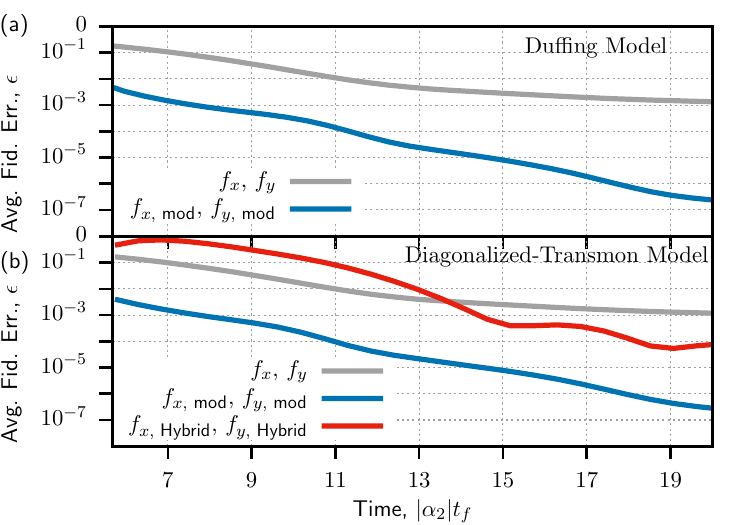}
    \caption{
        Model dependence of correction-pulse performance. Average fidelity error $\epsilon$ as a function of the dimensionless
        gate time $|\alpha_2| t_\uf$. (a) Results calculated with the Duffing Hamiltonian. The gray trace shows the calibrated
        baseline protocol, while the blue trace shows the self-consistent corrected protocol obtained using modified quadrature
        fields and a constant detuning. (b) Results evaluated with the diagonalized-transmon Hamiltonian. The gray and blue traces
        show the calibrated baseline and self-consistent corrected protocols, respectively. The red-orange trace shows the hybrid
        protocol, in which the Duffing-derived correction quadratures and detuning are combined with the independently calibrated
        diagonalized-transmon baseline pulse. The hybrid protocol is evaluated at the same physical gate time $t_\uf$ and performs
        substantially worse than the self-consistent diagonalized-transmon correction.
    }
    \label{fig:Tfig4_DuffVSFull}
\end{figure}

We now apply the correction procedure described above to the two four-level effective Hamiltonians introduced in
Sec.~\ref{sec:EffectiveH}. In both cases, the Hilbert space is truncated to the four lowest energy levels, as justified by the
truncation analysis of Fig.~\ref{fig:FidelityOfDifferentDimensionality}. For each model, we construct a modified Hamiltonian of
the form given in Eq.~\eqref{eq:Hmod}, and determine the correction quadrature envelopes $g_x (t)$ and $g_y (t)$, together with
the constant detuning $\Delta$ set by the drive frequency $\omega_\ud$, by imposing the Magnus-cancellation condition defined in
Eq.~\eqref{eq:MagnusCond}. The results shown in Fig.~\ref{fig:Tfig4_DuffVSFull} are obtained by keeping terms up to second order
in the Magnus expansion [$m=2$ in Eq.~\eqref{eq:MagnusCond}]. Applying this procedure to $\hH_\mm{diag}(t)$ gives the correction
Hamiltonian $\hW_\mm{diag}(t)$, while applying the same procedure to $\hH_\mm{Duff}(t)$ gives $\hW_\mm{Duff}(t)$.

Figure~\ref{fig:Tfig4_DuffVSFull} compares the resulting average fidelity error [see Eq.~\eqref{eq:fid}] as a function of the
dimensionless gate time $\abs{\alpha_2} t_\uf$. Figure~\ref{fig:Tfig4_DuffVSFull}(a) shows the results obtained with the Duffing
Hamiltonian. The gray trace corresponds to the baseline protocol, in which the uncorrected quadrature envelopes $f_x (t)$ and $f_y
(t) =0$ are used and the drive is resonant with the qubit transition. The blue trace shows the result of applying the Magnus-based
correction procedure using the Duffing Hamiltonian, yielding modified control fields $f_{x,\mm{mod}} (t)$ and $f_{y,\mm{mod}} (t)$
together with a constant detuning. The large reduction in the average fidelity error shows that, when the same Duffing Hamiltonian
is used both to construct and to evaluate the corrected protocol, the Magnus-based construction produces substantially improved
gates.

Figure~\ref{fig:Tfig4_DuffVSFull}(b) shows the corresponding comparison for the diagonalized-transmon model. The gray trace is
again the baseline protocol, now evaluated using $\hH_\mm{diag}(t)$. The blue trace shows the self-consistent corrected
protocol:~the modified control fields and constant detuning are computed using $\hH_\mm{diag} (t)$ and then evaluated with the same
diagonalized-transmon Hamiltonian. As in the Duffing case, the self-consistent correction substantially reduces the average
fidelity error over the gate-time range shown.

The red-orange trace in Fig.~\ref{fig:Tfig4_DuffVSFull}(b) represents the hybrid procedure. We first calculate the correction
envelopes $g_{\mm{Duff},x} (t)$ and $g_{\mm{Duff},y} (t)$, together with the constant detuning $\Delta_\mm{Duff}$, using the
Duffing Hamiltonian. In this calculation, the Duffing baseline envelope $f_x (t)$ is normalized using the Duffing charge-operator
matrix element $n_{0,1}$. To evaluate the predicted correction with the diagonalized-transmon Hamiltonian, we then add the
Duffing-derived correction envelopes to the independently calibrated diagonalized-transmon baseline pulse:
\begin{equation}
    \begin{aligned}
        f_{x,\mm{hybrid}} (t) &= f_{x,\mm{diag}} (t) + g_{x,\mm{Duff}} (t),\\
        f_{y,\mm{hybrid}} (t) &= g_{y,\mm{Duff}} (t),
    \end{aligned}
    \label{eq:HybridPulse}
\end{equation}
Here, $f_{x,\rm diag}(t)$ is normalized using the diagonalized-transmon value of $n_{0,1}$, while the Duffing-derived detuning
$\Delta_\mm{Duff}$ is used without further adjustment. Thus, the hybrid protocol transfers only the correction fields and
detuning, not the Duffing-calibrated baseline pulse. This construction tests whether the Duffing Hamiltonian correctly predicts
the additional controls required to correct an independently calibrated primitive gate.

For this comparison, the Duffing-derived correction is evaluated in the diagonalized-transmon model at the same physical gate time
$t_\uf$, not at the same value of $\abs{\alpha_2} t_\uf$. This distinction is important because the Duffing and diagonalized-transmon
models generally assign different anharmonicities to the leakage transition. Matching by $\abs{\alpha_2} t_\uf$  would therefore
compare corrections designed for different physical gate durations rather than testing whether the Duffing-predicted correction
transfers to the diagonalized-transmon dynamics at a fixed gate time.

The hybrid procedure provides a direct test of whether the correction predicted by the Duffing approximation remains effective
when combined with an independently calibrated baseline pulse and evaluated using the diagonalized-transmon Hamiltonian.
Figure~\ref{fig:Tfig4_DuffVSFull}(b) shows that this transfer is inefficient for the gate times considered here. At longer gate
times, the Duffing-derived correction can still reduce the average fidelity error relative to the baseline diagonalized-transmon
evolution, indicating that it captures part of the unwanted dynamics. However, its performance remains substantially worse than
the self-consistent diagonalized-transmon correction. At short gate times, the discrepancy is more severe:~The hybrid protocol can
fail to improve over the baseline protocol and may even increase the average fidelity error.

Taken together, the two panels of Fig.~\ref{fig:Tfig4_DuffVSFull} show that the error generator predicted by the Duffing
Hamiltonian is not interchangeable with that of the diagonalized-transmon Hamiltonian. Because the baseline pulse is independently
calibrated in the model used for evaluation, the degraded hybrid performance reflects a mismatch in the predicted multilevel
correction rather than a mismatch in the nominal qubit rotation. We next illustrate one manifestation of this dynamical model
dependence by comparing the drive-induced phase accumulation predicted by the two models.

\begin{figure}[t]
    \includegraphics[width=0.99\columnwidth]{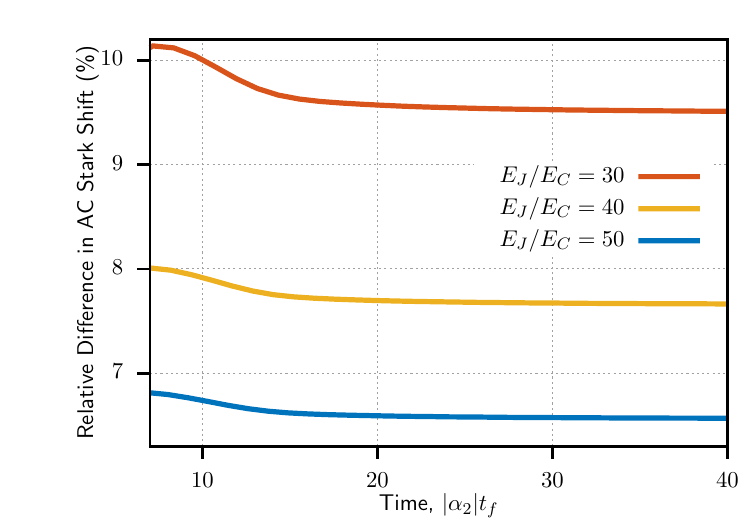}
    \caption{
    Model-dependent AC Stark phase mismatch. Relative difference, in percent, between the time-averaged AC Stark shifts predicted
    by the diagonalized-transmon and Duffing Hamiltonians as a function of dimensionless gate time $\abs{\alpha_2} t_\uf$. Equivalently,
    the plotted quantity measures the relative error in the net drive-induced phase accumulated by the qubit during the gate. The
    blue, yellow, and orange traces correspond to $E_J/E_C=50$, $40$, and $30$, respectively. The relative difference is largest
    in the fast-gate regime and increases as $E_J/E_C$ is reduced, illustrating how small differences in the model spectrum and
    charge-drive matrix elements can compound during driven evolution. The phase mismatch provides a diagnostic of dynamical model
    dependence.
    }
    \label{fig:ACStarkShiftFullvsDuff3x3}
\end{figure}

To illustrate how small differences between the two effective Hamiltonians can compound during driven evolution, we compare in
Fig.~\ref{fig:ACStarkShiftFullvsDuff3x3} the time-averaged AC Stark shifts predicted by the two models.  We emphasize that this is
not the instantaneous AC Stark shift during the pulse. Instead, it is the effective static shift associated with the net
drive-induced phase accumulated over the full gate. For the uncorrected Hamiltonians $\hH_\mm{diag}(t)$ and $\hH_\mm{Duff}(t)$, we
extract this phase from the diagonal part of the corresponding Magnus generator,
\begin{equation}
    \varphi_\mm{AC}^{(m)} (t_\uf) = \frac{i}{2} \left( \bra{1} \sum_{l=1}^{m} \hOmega_l (t_\uf) \ket{1} 
    - \bra{0} \sum_{l=1}^{m} \hOmega (t_\uf) \ket{0}\right),
    \label{eq:ACPhaseMagnus}
\end{equation}
where the sum is truncated at order $m$. The corresponding time-averaged AC Stark shift is then defined as
\begin{equation}
    \overline{\delta\omega}_\mm{AC}^{(m)} = \frac{\varphi_\mm{AC}^{(m)}(t_\uf)}{t_\uf}.
    \label{eq:TimeAveragedACStarkMagnus}
\end{equation}
This is the relevant quantity for the present control framework because the Magnus-based correction imposes conditions on the
final propagator rather than cancelling the error Hamiltonian at each instant in time. Consequently, the two models predict
different phase contributions to the error generators used to determine the correction envelopes. A correction constructed using
the time-averaged phase accumulation predicted by one effective Hamiltonian will not, in general, reproduce the phase compensation
predicted by the other. We use this mismatch as an illustrative diagnostic of how small, individually justified model
approximations can compound during driven evolution, rather than as a complete decomposition of the hybrid gate error.

Figure~\ref{fig:ACStarkShiftFullvsDuff3x3} shows the relative difference up to fourth order in the Magnus expansion, in percent,
between the time-averaged AC Stark shifts predicted by the diagonalized-transmon and Duffing Hamiltonians for $E_J/E_C=50$, $40$,
and $30$. For all three values, the mismatch is largest in the fast-gate regime and decreases for longer gates as the required
drive amplitude becomes smaller. The dependence on $E_J/E_C$ provides a diagnostic of the Duffing approximation. The relative
difference is smallest for $E_J/E_C=50$ and increases as $E_J/E_C$ is reduced, consistent with the device moving farther from the
deep-transmon regime where the quartic Duffing approximation and harmonic-oscillator charge matrix elements are most accurate.
Thus, individually small differences in the static spectrum and charge matrix elements can combine during driven evolution to
produce a sizable mismatch in the accumulated AC Stark phase.

This observation provides one illustration of the dynamical model dependence exposed by the hybrid protocol in
Fig.~\ref{fig:Tfig4_DuffVSFull}(b). The correction fields and detuning calculated with $\hH_\mm{Duff}(t)$ are constructed from the
effective error generator predicted by the Duffing Hamiltonian. When these correction fields are added to the independently
calibrated diagonalized-transmon baseline pulse, the spectrum and charge-drive matrix elements of $H_\mm{diag} (t)$ generate a
different driven response. The AC Stark-phase mismatch shown in Fig.~\ref{fig:ACStarkShiftFullvsDuff3x3} is one directly
identifiable manifestation of this difference, but it does not constitute a complete error budget. More generally, the result
demonstrates that approximations that appear accurate for static low-energy quantities need not preserve the dynamical error
generator relevant to predictive pulse design. The discrepancy becomes more visible as $E_J/E_C$ is reduced and the system moves
away from the deep-transmon regime.

\section{Validity of the Hilbert-space truncation for corrected dynamics}
\label{sec:ValidityTruncWithW}

In Sec.~\ref{sec:EffectiveH}, we introduced a criterion for truncating the Hilbert space of the driven transmon [see
Eq.~\eqref{eq:CutOffCond}]. That analysis was performed for the uncorrected Hamiltonian, before adding the correction Hamiltonian
$\hW(t)$. It is therefore important to verify that the same truncation remains valid for the modified Hamiltonian
$\hH_\mm{mod}(t)$ used to generate more accurate gates. Although one expects the correction pulse to reduce leakage out of the
computational subspace, this conclusion is not automatic. The correction is obtained perturbatively from a Magnus expansion, and
at sufficiently short gate times the larger drive amplitudes could, in principle, activate additional leakage channels that are
not accurately captured by a four-level model.

We test this explicitly using the diagonalized transmon Hamiltonian $\hH_\mm{diag}(t)$. We denote by
$\hW_{\mm{diag},N=4}(t)$ and $\hW_{\mm{diag},N=10}(t)$ the correction Hamiltonians computed from the
$N=4$ and $N=10$ truncations of $\hH_\mm{diag}(t)$, respectively. In both cases, the correction is obtained using
the same second-order Magnus-cancellation procedure. We then evaluate the corrected dynamics in both truncated Hilbert
spaces. This comparison directly tests whether a pulse derived from the four-level model remains valid when additional
leakage states are included.

\begin{figure}[t]
    \includegraphics[width=0.99\columnwidth]{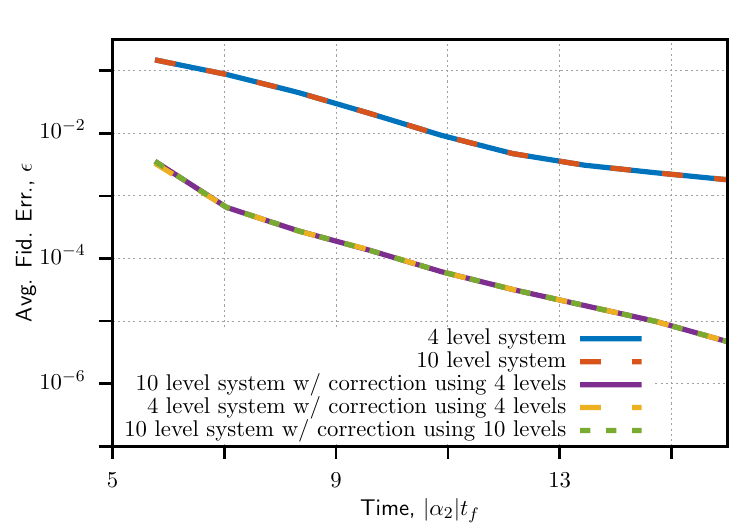}
    \caption{
        Validity of the four-level truncation for corrected dynamics. Average fidelity error for the diagonalized-transmon
        Hamiltonian using four- and ten-level Hilbert-space truncations. The uncorrected four- and ten-level evolutions agree
        over the gate-time range shown, confirming the convergence observed in Fig.~\ref{fig:FidelityOfDifferentDimensionality}.
        After correction, a pulse computed in the four-level model gives essentially the same performance when applied to the
        ten-level model as a pulse computed self-consistently in the ten-level model. Thus, for the gate times considered here,
        the four-level truncation captures the relevant leakage channels not only for the uncorrected dynamics but also for the
        corrected evolution.
    }
    \label{fig:4vs10LevelsW}
\end{figure}

The results are shown in Fig.~\ref{fig:4vs10LevelsW}. As discussed in connection with
Fig.~\ref{fig:FidelityOfDifferentDimensionality}, increasing the truncation of $\hH_\mm{diag}(t)$ from $N=4$ to $N=8$ does not
lead to appreciable quantitative differences in the average fidelity error. In Fig.~\ref{fig:4vs10LevelsW}, the solid blue and
dashed orange traces show the corresponding dynamics generated by $\hH_\mm{diag}(t)$ without adding $\hW(t)$ for $N=4$ and $N=10$,
respectively. These traces are included only as baselines for comparison. The central question is whether the same truncation
remains valid after adding the correction Hamiltonian. The dashed
yellow trace shows the result of applying $\hW_{\mm{diag},N=4}(t)$ to the $N=4$ truncation, while the dashed green
trace shows the result of applying $\hW_{\mm{diag},N=10}(t)$ to the $N=10$ truncation. These two traces show no
appreciable quantitative difference. Thus, computing the correction Hamiltonian in a four-level Hilbert space already
captures the corrected dynamics obtained from a ten-level calculation.

As a final and more stringent test, the solid purple trace shows the result of applying the modified envelopes obtained from
$\hW_{\mm{diag},N=4}(t)$ to the $N=10$ truncation of $\hH_\mm{diag}(t)$. This curve is essentially indistinguishable from the
self-consistent result obtained with $\hW_{\mm{diag},N=10}(t)$. For the gate times considered here, a pulse
derived from the four-level model remains valid when six additional leakage levels are included in the time evolution. The
truncation criterion of Sec.~\ref{sec:EffectiveH}, originally motivated using the uncorrected Hamiltonian, continues to hold for
the corrected dynamics generated by $\hH_\mm{mod}(t)$.

This validation is important for the rest of the analysis. It shows that the discrepancies discussed in
Fig.~\ref{fig:Tfig4_DuffVSFull} do not arise because the four-level model misses higher leakage states once the
correction pulse is applied. Rather, the dominant effect is the model dependence of the effective Hamiltonian itself.
We can therefore use four-level corrected Hamiltonians in the following analysis without introducing a significant
truncation error.

\section{Model dependence of the control framework}
\label{sec:LinearCorrectionBreakdown}

The preceding section showed that a four-level truncation is sufficient to describe the corrected dynamics generated by
$\hH_\mm{diag} (t)$ for the gate times considered here. We now show that the choice of truncation also affects a different aspect
of the control problem:~The structure of the error channels that the correction pulse must cancel. If the effective Hilbert space
is too small, some error channels are absent from the Hamiltonian used for pulse design. The model may then suggest that a
restricted correction framework is sufficient, even though a more faithful Hamiltonian reveals additional channels that must be
addressed.

More generally, the effective model should inform the choice of control framework. A simplified model may indicate that a
relatively simple correction strategy is sufficient, whereas a more accurate model can reveal additional error channels that
require a more flexible approach. We illustrate this point within the constrained-control framework introduced in
Sec.~\ref{sec:OLQC}. In this framework, the correction Hamiltonian $\hW (t)$ is constructed perturbatively. We refer to the
strategy as linear when, at each order of the expansion, the required correction $\hW^{(n)} (t)$ can be written using the
experimentally available lab-frame control parameters and can directly cancel the unwanted leakage transitions and phase errors at
that order. This construction succeeds when the error-canceling operators required by the model are compatible with the available
controls. When this is not the case, a nonlinear strategy is required. The missing effective operators are generated through
higher-order Magnus terms, including commutators involving $\hW (t)$ and the Hamiltonian terms of the uncorrected
dynamics~\cite{roque2021}. Thus, the model Hamiltonian does not merely determine the predicted fidelity of a given pulse; it also
determines which error channels are visible and therefore which correction framework is appropriate. In short, the linear strategy
uses correction operators that are directly available through the laboratory controls, whereas the nonlinear strategy relies on
higher-order commutators to generate additional effective operator directions.

This distinction becomes relevant when comparing three- and four-level descriptions of the transmon. In a three-level model, the
error channels exposed by the Hamiltonian can be canceled order by order using correction terms $\hW^{(n)}(t)$ that are directly
compatible with the available lab-frame control parameters. The linear strategy is therefore effective in this reduced
description. Once the fourth level is included, additional higher-order error channels appear that cannot be removed by such
directly implementable correction terms alone. Since the preceding sections showed that the four-level model captures the relevant
driven dynamics, this limitation reflects the physical control problem rather than an artifact of an unnecessarily large
truncation.

We illustrate this mechanism by comparing three- and four-level descriptions of the transmon. A three-level model contains a
single noncomputational state, $|2\rangle$, and therefore captures only the leakage channels involving the first leakage level. In
contrast, a four-level model also includes $|3\rangle$. This additional state enables higher-order processes involving virtual or
real excursions through the leakage subspace. Even when the final population in $|3\rangle$ remains small, such processes can
contribute to the accumulated phases of the computational states and modify the effective error generator. These contributions are
absent in a three-level truncation but are present in the four-level model.

\begin{figure}[t]
    \includegraphics[width=0.99\columnwidth]{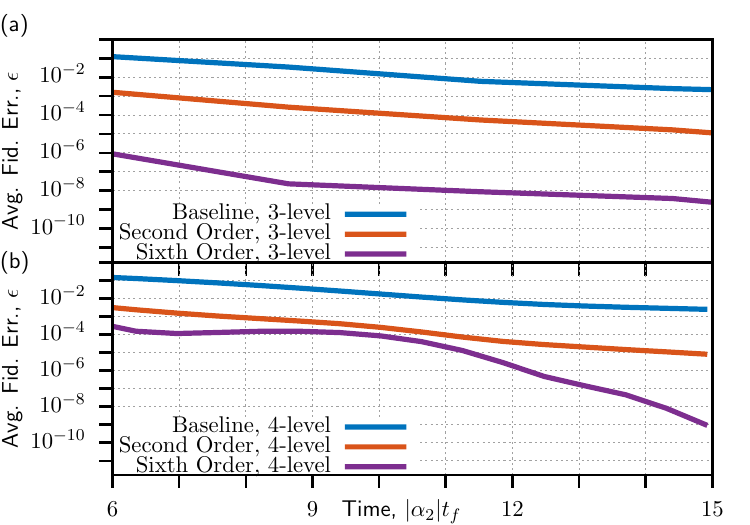}
    \caption{
        Effect of Hilbert-space truncation on the apparent performance of restricted correction strategies. Average fidelity error
        $\epsilon$ as a function of dimensionless gate time $\abs{\alpha_2} t_\uf$ for linear corrections constructed in the
        diagonalized-transmon model using (a) three levels and (b) four levels. In the three-level model, increasing the
        correction order from second to sixth order systematically improves the gate fidelity, indicating that the visible error
        channels can be addressed within the restricted correction framework. In the four-level model, the second-order correction
        still reduces the average fidelity error, but increasing to sixth order no longer yields comparable improvement, especially at short gate
        times. The additional level reveals higher-order error channels that are absent in the three-level truncation and that
        require a more flexible correction framework.
    }
    \label{fig:HigherOrdersLinearCombo}
\end{figure}

Figure~\ref{fig:HigherOrdersLinearCombo} demonstrates the consequence for linear corrections.
Figure~\ref{fig:HigherOrdersLinearCombo}(a) shows the average fidelity error for the three-level model. The uncorrected evolution
is shown as a baseline (blue trace), while the other traces correspond to linear corrections constructed by cancelling the Magnus
expansion at second (orange trace) and sixth order (purple trace). In this reduced model, increasing the correction order produces
the expected improvement:~The second-order correction lowers the error relative to the uncorrected pulse, and the sixth-order
correction lowers it further over the gate-time range shown. Thus, within the three-level model, the error channels generated by
the dynamics are compatible with the available charge-drive correction.

The behavior changes once the fourth level is retained, as shown in Fig.~\ref{fig:HigherOrdersLinearCombo}(b). The second-order
linear correction (orange trace) again reduces the average fidelity error relative to the uncorrected evolution (blue trace) and
gives results comparable to those obtained in the three-level model. However, increasing the correction order from second to sixth
order (purple trace) no longer leads to a systematic improvement. In particular, at short gate times the corrected error saturates
well above the value one would expect if all relevant higher-order errors could be cancelled. This saturation indicates that the
limitation is not simply the truncation order of the Magnus expansion. Rather, the four-level dynamics generates higher-order
error components that are not removed by the available linear correction.

The comparison between Figs.~\ref{fig:HigherOrdersLinearCombo}(a) and (b) therefore provides a useful diagnostic. In the
three-level model, the linear construction appears sufficient because the truncated Hilbert space omits channels involving
$\ket{3}$. Once those channels are included, increasing the linear correction order no longer yields systematic improvement. This
indicates that the limitation is not merely the order of the Magnus expansion, but the requirement that the linear correction
terms remain directly implementable using the available lab-frame control parameters.

This observation highlights a second role of the truncation analysis developed above. An effective Hamiltonian should both
reproduce the dynamics generated by a given pulse and expose the error structure relevant to the control problem. For the transmon
parameters considered here, the four-level model reveals error channels that limit the performance of linear corrections at fast
gate times. This motivates the extended correction framework considered next.

\section{Extended Corrections in the Four-Level Transmon Model}
\label{sec:NonLinCorr}

The results of Sec.~\ref{sec:LinearCorrectionBreakdown} show that the four-level model reveals error channels that are not
effectively mitigated by the linear correction strategy at short gate times. We now apply the nonlinear correction strategy within
the same constrained-control setting. The physical control resources are unchanged:~The corrected protocol uses the same
quadrature correction fields and the same constant detuning control as in the linear case. The purpose of this section is to show
that this extended framework provides systematic improvement beyond the linear construction and produces a genuinely different
control solution.

\begin{figure}[t]
    \includegraphics[width=9cm]{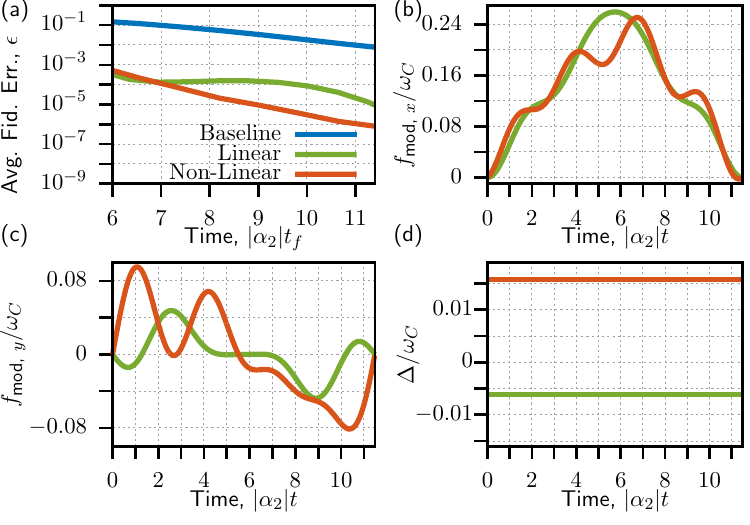}
    \caption{
        Extended correction framework in the four-level diagonalized-transmon model. (a) Average fidelity error $\epsilon$ as a
        function of dimensionless time $\abs{\alpha_2} t_\uf$ for the uncorrected and corrected gates obtained using fourth-order
        linear and nonlinear constructions. For $\abs{\alpha_2} t_\uf \gtrsim 7$, the nonlinear correction gives an additional
        reduction of roughly one order of magnitude compared with the linear correction, while the two fourth-order corrections
        give comparable errors at shorter gate times. (b)--(d) Fourth-order modified control fields for a representative gate
        time, showing the in-phase component $f_{\mm{mod},x}(t)$, quadrature component $f_{\mm{mod},y} (t)$, and detuning
        component $\Delta$, respectively. The linear and nonlinear correction strategies lead to visibly different pulse shapes,
        demonstrating that the improvement in panel (a) arises from a different control solution rather than from a small
        perturbative adjustment of the same pulse.
    }
    \label{fig:NewNonLinearMethod}
\end{figure}

Figure~\ref{fig:NewNonLinearMethod} compares fourth-order linear and nonlinear corrections in the four-level
diagonalized-transmon model. Figure~\ref{fig:NewNonLinearMethod}(a) shows the average fidelity error as a function of the
dimensionless gate time $\abs{\alpha_2} t_\uf$. The blue trace is the uncorrected baseline, while the green and orange traces
correspond to the fourth-order linear and nonlinear corrections, respectively. Both corrected protocols reduce the average
fidelity error relative to the uncorrected pulse. For $\abs{\alpha_2} t_\uf \gtrsim 7$, the nonlinear strategy yields an
additional reduction of roughly one order of magnitude compared with the linear strategy. At shorter gate times, however, the
two fourth-order corrections give comparable errors. Thus, the main significance of the nonlinear construction is not simply the
magnitude of the improvement at fourth order, but the fact that it cancels the relevant fourth-order error channels using the
available physical controls. This opens a systematic route to pushing the nonlinear correction to higher order, which is expected
to further reduce the error in the shortest-time regime. This behavior should be contrasted with the linear strategy in
Fig.~\ref{fig:HigherOrdersLinearCombo}(b), where increasing the correction order from second to sixth does not lead to a comparable
improvement for $\abs{\alpha_2}t_\uf \lesssim 10$.

Figures~\ref{fig:NewNonLinearMethod}(b) to (d) show that this improvement is accompanied by a qualitatively different control
solution. These panels compare the modified pulses obtained from the fourth-order linear and nonlinear constructions for a
representative gate time. In each panel, the resulting linear modified pulse is plotted in green, while the resulting nonlinear
modified pulse is plotted in orange. Figure~\ref{fig:NewNonLinearMethod}(b) shows the modified in-phase component
$f_{\mm{mod},x}(t)$, Fig.~\ref{fig:NewNonLinearMethod}(c) shows the modified quadrature component $f_{\mm{mod},y}(t)$, and
Fig.~\ref{fig:NewNonLinearMethod}(d) shows the detuning component $\Delta$. All three quantities are plotted in units of
$\omega_\uC$. The linear and nonlinear frameworks lead to visibly different time-dependent controls in all three components. Thus,
the improvement in Fig.~\ref{fig:NewNonLinearMethod}(a) is not merely the result of a small perturbative adjustment to the same
pulse. Rather, the nonlinear strategy identifies a distinct pulse that more effectively compensates the error channels present in
the four-level model.

This comparison demonstrates that the limited performance of the fourth-order linear correction is not due to a lack of physical
control resources. The nonlinear correction uses the same available controls, but combines them differently and thereby reaches a
more effective solution. Together with Secs.~\ref{sec:ValidityTruncWithW} and \ref{sec:LinearCorrectionBreakdown},
Fig.~\ref{fig:NewNonLinearMethod} completes the logic of the analysis:~The four-level model captures the relevant corrected
dynamics, reveals limitations of the linear strategy that are hidden in a three-level truncation, and motivates an extended
framework capable of producing genuinely different pulse shapes. Thus, the effective Hamiltonian does more than determine the
numerical accuracy of a simulation:~It also identifies the error channels that must be addressed and thereby informs the
appropriate control framework for high-fidelity gates.

\section{Conclusion}

We have shown that the effective Hamiltonian used for pulse design can qualitatively affect the performance of open-loop quantum
control. For fast single-qubit gates in a capacitively driven transmon, we compared correction protocols constructed using the
standard Duffing approximation with protocols constructed from a Hamiltonian obtained by diagonalizing the static transmon
Hamiltonian and expressing the physical drive in the resulting eigenbasis. Although Duffing-derived corrections can substantially
reduce the gate error predicted by the Duffing Hamiltonian, their performance is substantially degraded when the correction fields
are added to an independently calibrated baseline pulse in the diagonalized-transmon model. The nominal qubit rotation is
calibrated separately in each model through the corresponding value of $n_{0,1}$, so the discrepancy reflects the predicted
multilevel correction rather than a trivial mismatch in the baseline Rabi rate.

The Duffing approximation introduces small and individually well-justified changes in both the transition frequencies and the
charge-drive matrix elements. Our results show that the apparent accuracy of these approximations at the level of static
low-energy quantities does not guarantee accurate predictions under driven evolution. The differences can compound over the
duration of a control pulse and lead the two Hamiltonians to predict different dynamical error generators and hence different
correction protocols. The mismatch in the accumulated AC Stark phase provides one illustrative diagnostic of this effect, rather
than a complete decomposition of the total gate error. The discrepancy becomes more visible as $E_J/E_C$ is reduced, emphasizing
the importance of evaluating an effective Hamiltonian according to the driven control task for which it will be used. 

We further showed that the model Hamiltonian also informs the choice of control framework. A four-level diagonalized-transmon
model is sufficient to reproduce the corrected dynamics obtained with larger Hilbert-space truncations, but a three-level model
removes higher-order error channels involving the second leakage level. As a result, the three-level model can make a linear
correction strategy appear sufficient, whereas the four-level model reveals limitations of that strategy. Using an extended
nonlinear correction framework provides a systematic route to high-fidelity performance while relying on the same physical control
resources, and produces pulse shapes that differ substantially from those obtained with the linear construction.

These results demonstrate that simplified Hamiltonians can lead to inefficient or misleading control pulses, even for single-qubit
gates. This conclusion concerns coherent model dependence under an ideal applied control waveform. Decoherence, waveform
distortion, parameter drift, and control-line uncertainty introduce additional effects and can be incorporated into an extended
control model, but they are not required to expose the Hamiltonian-dependent discrepancy studied here.  The model used for pulse
design must therefore be chosen not only to reproduce spectra or uncorrected dynamics, but also to expose the error channels
relevant to the intended control task. Model selection is thus an essential component of high-fidelity quantum-control design.

\appendix

\section{Derivation and validation of the sequential-transition estimate}
\label{app:Omega1MatrixElements}

In this Appendix, we derive the sequential-transition estimate $\eta_k$ introduced in Eq.~\eqref{eq:CutOffCond} and
validate its use in determining the Hilbert-space truncation. Starting from the interaction-picture Hamiltonian  $\hV_\uI (t)$ in
Eq.~\eqref{eq:HrotIntPict}, we evaluate the matrix elements of the first-order Magnus term at $t=t_\uf$. Their squared magnitudes
give the elementary transition probabilities defined in Eq.~\eqref{eq:1stOrdMagnusProb}. Since $\hV_\uI (t)$ couples
predominantly adjacent transmon levels, the sequential-transition estimate for reaching a higher level is constructed by
multiplying the elementary first-order probabilities along the corresponding adjacent-level pathway.

For the choice of uncorrected initial pulse, i.e., $f_y(t)=0$ and $f_x(t)$ given by Eq.~\eqref{eq:fx}, the envelope satisfies
boundary conditions that suppress off-resonant transitions. In particular, $f_x (0) = f_x (t_\uf) = 0$ and $\dot{f}_x (0) =
\dot{f}_x (t_\uf) = 0$. Repeated integration by parts therefore removes the first two boundary contributions, and the leading
contribution to an off-resonant transition amplitude scales as $(\alpha_k t_\uf)^{-3}$.  More generally, we denote by $s$ the
smoothness exponent associated with this boundary cancellation; for the pulse in Eq.~\eqref{eq:fx} we have $s=3$.

The first leakage channel couples the computational subspace to $\ket{2}$. From Eq.~\eqref{eq:HrotIntPict}, we find that the
first-order Magnus transition amplitudes at $t=t_\uf$ are
\begin{equation}
    \begin{aligned}
       \bra{0} \hOmega_1 (t_\uf)\ket{2} &= -i \int_0^{t_\uf} \di{t} \bra{0} \hV_\uI (t) \ket{2} \\
           &= -\frac{n_{1,2}}{2} \int_0^{t_\uf} \di{t} e^{i \alpha_2 t} f_x (t) \sin\left[ \frac{\theta (t)}{2}\right] \\
           &= -2 i \frac{n_{1,2}}{n_{0,1}} \frac{\pi^2 \theta_0}{(\alpha_2 t_\uf)^3} \sin\left(\frac{\theta_0}{2}\right) e^{i
           \alpha_2 t_\uf} \\
           &\phantom{={}} +\mathcal{O}\left(\frac{1}{(\alpha_2 t_\uf)^4}\right),
    \end{aligned}
	\label{eq:Omega1_02}
\end{equation}
where we used integration by parts to obtain the third equality. Proceeding similarly, we obtain
\begin{equation}
    \begin{aligned}
       \bra{1} \hOmega_1 (t_\uf)\ket{2} &= -i \int_0^{t_\uf} \di{t} \bra{1} \hV_\uI (t) \ket{2} \\
           &= -i \frac{n_{1,2}}{2} \int_0^{t_\uf} \di{t} e^{i \alpha_2 t} f_x (t) \cos\left[ \frac{\theta (t)}{2}\right] \\
           &= 2 \frac{n_{1,2}}{n_{0,1}} \frac{\pi^2 \theta_0}{(\alpha_2 t_\uf)^3} \left(-1 +  \cos\left(\frac{\theta_0}{2}\right)
           e^{i \alpha_2 t_\uf}\right) \\
           &\phantom{={}} +\mathcal{O}\left(\frac{1}{(\alpha_2 t_\uf)^4}\right).
    \end{aligned}
	\label{eq:Omega1_12}
\end{equation}

The first-order Magnus transition amplitude at $t=t_\uf$ between neighboring leakage states $\ket{k}$ and $\ket{k+1}$ ($k\geq 2$) is 
\begin{equation}
    \begin{aligned}
       \bra{k} \hOmega_1 (t_\uf)\ket{k+1} &= -i \int_0^{t_\uf} \di{t} \bra{k} \hV_\uI (t) \ket{k+1} \\
           &= -i \frac{n_{k,k+1}}{2} \int_0^{t_\uf} \di{t} e^{i \alpha_{k+1} t} f_x (t)  \\
           &= -2 \frac{n_{k,k+1}}{n_{0,1}} \frac{\pi^2 \theta_0 \left( e^{i \alpha_{k+1} t_\uf} -1 \right)}{\alpha_{k+1}t_\uf \left[4 \pi^2 - \left(\alpha_{k+1} t_\uf
           \right)^2\right]}.
    \end{aligned}
	\label{eq:Omega1_kkp1}
\end{equation}

Using the definition in Eq.~\eqref{eq:1stOrdMagnusProb}, the amplitudes in Eqs.~\eqref{eq:Omega1_02}, \eqref{eq:Omega1_12}, and
\eqref{eq:Omega1_kkp1} yield the corresponding first-order transition probabilities. For transitions from the computational
subspace to $\ket{2}$, we obtain
\begin{equation}
    \begin{aligned}
        P^{(1)}_{\ket{0} \to \ket{2}} &= \abs{\bra{0} \hOmega_1 (t_\uf)\ket{2}}^2 \\
        &\simeq 4 \frac{n_{1,2}^2}{n_{0,1}^2} \frac{\pi^4 \theta_0^2 \sin^2\left(\frac{\theta_0}{2}\right)}{\abs{\alpha_2 t_\uf}^6} \\
        &\leq 4 \frac{n_{1,2}^2}{n_{0,1}^2} \frac{\pi^4 \theta_0^2}{\abs{\alpha_2 t_\uf}^6}
    \end{aligned}
    \label{eq:P1_02}
\end{equation}
and
\begin{equation}
    \begin{aligned}
        P^{(1)}_{\ket{1} \to \ket{2}} &= \abs{\bra{1} \hOmega_1 (t_\uf)\ket{2}}^2 \\
        &\simeq 2 \frac{n_{1,2}^2}{n_{0,1}^2} \frac{\pi^4 \theta_0^2 \left[3 - 4 \cos\left( \alpha_2 t_\uf  \right)
        \cos\left(\frac{\theta_0}{2}\right) + \cos\left(\theta_0\right)\right]}{\abs{\alpha_2
        t_\uf}^6} \\
        &\leq 16 \frac{n_{1,2}^2}{n_{0,1}^2} \frac{\pi^4 \theta_0^2 \cos^4\left(\frac{\theta_0}{2}\right)}{\abs{\alpha_2 t_\uf}^6} \\
        &\leq 16 \frac{n_{1,2}^2}{n_{0,1}^2} \frac{\pi^4 \theta_0^2}{\abs{\alpha_2 t_\uf}^6},
    \end{aligned}
    \label{eq:P1_12}
\end{equation}
where, for the purpose of deriving a truncation criterion, we have bounded $P^{(1)}_{\ket{1} \to \ket{2}}$ in the last line.
For transitions between adjacent leakage states $\ket{k}$ and $\ket{k+1}$, with $k\geq2$, Eq.~\eqref{eq:Omega1_kkp1} similarly
gives
\begin{equation}
    \begin{aligned}
        P^{(1)}_{\ket{k} \to \ket{k+1}} &= \abs{\bra{k} \hOmega_1 (t_\uf)\ket{k+1}}^2 \\
            &\simeq 8 \frac{n_{k,k+1}^2}{n_{0,1}^2}   
            \frac{\pi^4 \theta_0^2 \left[ 1- \cos\left(\alpha_{k+1} t_\uf \right) \right]}{\left(\alpha_{k+1} t_\uf\right)^2 \left[-4\pi^2 + \left(\alpha_{k+1} t_\uf\right)^2\right]^2} \\
            &\lesssim 16  \frac{n_{k,k+1}^2}{n_{0,1}^2} 
            \frac{\pi^4 \theta_0^2}{\left(\alpha_{k+1} t_\uf\right)^2 \left[-4\pi^2 + \left(\alpha_{k+1} t_\uf\right)^2\right]^2},
    \end{aligned}
    \label{eq:P1_kkp1}
\end{equation}
where, similarly to our treatment of $P^{(1)}_{\ket{1} \to \ket{2}}$, we have bounded $P^{(1)}_{\ket{k} \to \ket{k+1}}$ in the last
line since our goal is to obtain a truncation criterion.

\begin{figure}[t]
    \includegraphics[width=0.99\columnwidth]{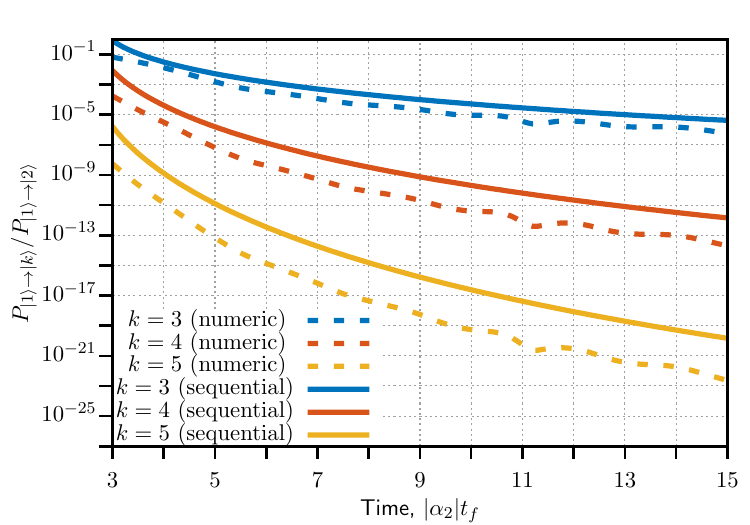}
	\caption{
        Validation of the leakage hierarchy underlying Eq.~\eqref{eq:CutOffCond}. The plot compares normalized higher-level
        leakage quantities for pathways originating from $\ket{1}$. The curves labeled ``sequential'' show the corresponding
        normalized sequential-transition estimates, whereas the curves labeled ``numeric'' show the normalized transition
        probabilities $P_{\ket{1}\to\ket{k}}/P_{\ket{1}\to\ket{2}}$ obtained by numerically solving Schrödinger's equation. The
        comparison shows that the sequential-transition estimate captures the hierarchy and gate-time scaling of the leading
        higher-level leakage channels over the range considered.
    }
    \label{fig:appLeakageAmplitudes}
\end{figure}

To rank the successive leakage channels, we define the sequential-transition estimate for a pathway originating from a
computational state $\ket{l}$, $l=0,1$, and reaching a leakage state $\ket{k}$, $k\geq2$, by multiplying the corresponding
adjacent first-order Magnus transition probabilities:
\begin{equation}
    \eta_{l \to k} \equiv P^{(1)}_{\ket{l} \to \ket{2}} \prod_{m=3}^k P^{(1)}_{m-1,m}.
    \label{eq:EstTransProb}
\end{equation}
Equation~\eqref{eq:EstTransProb} is not an exact perturbative transition probability. In particular, it does not retain the phases
and interference terms appearing in a complete higher-order calculation. Its purpose is to capture the parametric hierarchy of the
leakage channels and thereby identify the level at which further extension of the Hilbert space becomes numerically negligible.

Figure~\ref{fig:appLeakageAmplitudes} validates the hierarchy implied by Eq.~\eqref{eq:EstTransProb}. The figure compares the
normalized sequential-transition estimates with the corresponding normalized transition probabilities obtained from direct
numerical integration of Schrödinger's equation. The agreement for the leading higher-leakage channels shows that products of
adjacent first-order probabilities capture the dominant hierarchy and gate-time scaling. Moreover, the channels reaching $\ket{4}$
and higher levels are strongly suppressed relative to the channel reaching $\ket{3}$. This confirms that, for the gate-time window
considered in the main text, the four-level truncation retains the dominant leakage channel beyond $\ket{2}$ while neglecting only
parametrically smaller contributions.

As shown by the upper-bound estimates in Eqs.~\eqref{eq:P1_02} and \eqref{eq:P1_12}, the $\ket{1}\to\ket{2}$ estimate is larger
than the $\ket{0}\to\ket{2}$ estimate by a factor of $4$. We therefore use the pathway starting from $\ket{1}$ to define the
conservative sequential-transition estimate $\eta_k\equiv\eta_{1\to k}$. Substituting the bounds in Eqs.~\eqref{eq:P1_12} and
\eqref{eq:P1_kkp1} into Eq.~\eqref{eq:EstTransProb} gives the expression for $\eta_k$ quoted in Eq.~\eqref{eq:CutOffCond}.

\bibliography{QuantumCtrlModels.bib}

\end{document}